**Sample Size and Bias Approximations For Continuous Exposures Measured with Error**

Honghyok Kim

Division of Environmental and Occupational Health Sciences, School of Public Health, University of Illinois Chicago, Chicago, IL, 60661, USA

Research and Management Center for Health Risk of Particulate Matter, Seoul, 02841, Republic of Korea

honghyok@uic.edu

Supplementary materials: p31 to p70

**Abstract**

Measurement error is a pervasive challenge across many disciplines, yet its impact on sample size determination and the accuracy and precision of estimators regarding the association between an exposure and an outcome remains understudied in real-world complex scenarios. These include heteroskedastic continuous exposures, error-prone measurements, multiple exposure time points, and the use of calibrated exposure variables. This article develops approximation equations for sample size calculations, estimator accuracy, and standard errors of the estimator in estimating the effect of an exposure on an outcome. For sample size calculations, as an example, we focus on (nested) matched case-control studies with conditional logistic regression. But they could be extended to other settings with sample size equations elsewhere. Our approximation of estimator accuracy is based on linear model approximations that can be applied to logistic regression and linear models. This paper considers non-linear effect estimation using polynomials and addresses non-differential, autocorrelated, and differential additive or multiplicative measurement errors in distributed lag models for heteroskedastic exposures in the absence or presence of exposure validation data. The proposed framework will provide insights into efficient research design and a deeper understanding of measurement error impacts on research.

# 1. Introduction

Measurement error has long been recognized as a pervasive challenge, yet significant gaps remain in understanding its implications for sample size determination and the accuracy and precision of estimators. These challenges are particularly acute in real-world complex scenarios involving heteroskedastic exposures, distributed lag (non-linear) models, and various measurement error structures. Addressing these issues will help ensure robust causal inference, evidence synthesis, and efficient study design (Brakenhoff et al. , 2018, Innes et al. , 2021, Loken & Gelman, 2017).

The heteroskedastic variance of continuous exposures is prevalent, such as time-varying exposure. In addition, distributed lag (non-linear) models are often employed to identify critical time-windows—such as cumulative exposure, lag time, or induction periods—or to isolate effects at specific time points (Danieli et al. , 2020, Ganjavi & Faraji, 2019, Gasparrini et al. , 2010, Kaier et al. , 2009, Kim & Lee, 2019, Mork et al. , 2024). These issues are further complicated by measurement error structures that affect the accuracy and precision of estimators and sample size determination. Another layer of complexity arises from residual confounding due to covariate measurement error, particularly when other time periods of the exposure act as confounders. This can lead to biased estimates of critical time-windows or specific exposure effects (Fewell et al. , 2007, Greenland & Robins, 1985, Kim & Lee, 2019). Moreover, investigators may often lack direct access to validation data for exposure measurement. They may instead rely on summary statistics, such as coefficients of determination or regression coefficients, yet how to leverage this information to predict estimator accuracy and precision, or determine sample size, remains unclear. Even when validation studies are feasible, integrating corrected exposure variables into sample size calculations for the main study is methodologically complex, particularly in scenarios involving heteroskedastic distributed lag models.

To address these gaps, this paper develops approximation equations for sample size, estimator bias, and standard errors under heteroskedastic distributed lag exposure models, with and without measurement errors. For scenarios with validation data, this paper proposes highly accurate approximation equations. When validation data is unavailable, this paper provides practical, albeit less precise, alternatives based on summary accuracy measures. Although sample size equations in this paper are grounded in conditional logistic regression with matched samples, but the framework can be extended to other settings, such as unconditional logistic regression (via case incidence adjustment) and linear regression for continuous outcomes (via outcome variance adjustment). Bias approximations are derived from linear models, ensuring broad applicability. This paper was built upon previous work including sample size equations in the context of continuous exposure, including conditional logistic regression (Lachin, 2008) and Berkson, classical, and linear measurement error in one continuous covariate, primarily within generalized linear model settings (Bernardo et al. , 2000, Keogh et al. , 2020, Tosteson et al. , 2003, White et al. , 1994). The present paper additionally considers heteroskedastic continuous exposure variables, distributed lags, complex measurement errors including autocorrelated errors and multiplicative errors, and how to approximate sample size and bias with and without validation data to offer a deeper understanding of measurement errors on research and solutions for designing robust studies under real-world constraints.

2. **Motivating examples**

This paper was motivated by conditional logistic regression in (nested) case-control studies with one-to-k matching, which are widely used in many different fields. Specifically, as demonstrated in the application section below, a motivating example is to understand the actual sample size in a prospective or retrospective birth cohort study investigating the effect of prenatal exposure on health outcomes in infants or children. There are other study designs in various disciplines that may be seen as a variation of case-control designs because *case* means an individual (or group depending on the level of analysis) having the occurrence of a binary outcome. Matching can be conducted using important confounder or case/control selection-related variables. Propensity score matching may be used. Also, self-controls may be used. For example, case-crossover designs are matched case-control designs that utilize information from the comparison between the time of each incident (i.e., index time) and self-control times selected from pre-specified time points to estimate causal effects (Lee & Schwartz, 1999, Maclure, 1991, Mittleman & Mostofsky, 2014).

3. **Notations**

Let $x_{i,t}$ denote the unmeasured true exposure of interest, $x$, for a case or (self-)control $i$ at day $t$. Let $x_{i,t-l}$ denote the temporally *l*-lagged variable of $x_{i,t}$, called *Lagl* exposure. Let $x_{g,t}$ denote the exposure variable measured or estimated at spatial grid/area $g$ at day $t$. Sometimes, $x_{g,t}$ is used in place of $x_{i,t}$ (e.g., measurements from environmental monitors or area-level exposure estimates). If not noted, I suppose that $g$ is spatially less granular than case/(self-)control $i$'s location and $g$ subsumes $i$'s location. $\boldsymbol{x_{i\ or\ g}} = (x_{i\ or\ g,t}, \ldots, x_{i\ or\ g,t-L})$ and $\boldsymbol{x_{i\ or\ g}^{-l}} = (x_{i\ or\ g,t}, \ldots, x_{i\ or\ g,t-(l-1)}, x_{i\ or\ g,t-(l+1)}, \ldots, x_{i\ or\ g,t-L})$ may be used, where $L$ denotes the maximum lag. Let $x_{t-l}$ denote $x_{i\ or\ g,t-l}$. I use $x_{t-l}$ for when there does not exist exposure measurement error at all and the subscript $i$ or $g$ does not contain any additional information. For example, if $x_{i,t-l}$ is an actual exposure but $x_{g,t-l}$ is used for effect estimation, then $x_{g,t-l}$ may be an error-prone variable because $g$ is spatially less granular. I may need to distinguish these two by specifically noting $x_{i,t-l}$ or $x_{g,t-l}$. If $x_{i,t-l}$ is used for effect estimation, then there is no need to use the subscript $i$ and I may use $x_{t-l}$. In some research setting (e.g., group-level inference), $x_{i,t-l} \cong x_{g,t-l}$, implying no specific need to use the subscript $i$ or $g$.

Let $Y_i$ denote the outcome of a (self-)control ($Y_i = 0$) or a case ($Y_i = 1$). Let $\mathbf{z}$ denote a set of covariates other than the main exposure variable(s) such as potential confounders, including dummy variables indicating each case-(self-)controls matched set. Assume $n$ cases independently originate from a source population given $x$ and $\mathbf{z}$, each matched to their respective (self-)controls. Each set has one case and $m_h - 1$ (self-)controls, resulting in a $m_h$-sized matched set, where $h$ indicates $h$-th case. Note that $n$ also indicates the number of sets because each set has only one case. $x_{h,j,l}$ indicates $x_{t-l}$ for $j$-th observation in $h$-th matched set. Let 1-st observation in each set be the case and the others be (self-)controls. Let $\sigma_h^2(v)$ denote the variance of a variable, $v$, in $h$-th matched set without "-1" correction. Note that superscript $^{ep}$ such as $x^{ep}$ is used to denote non-Berkson error-prone variable (e.g., a linear or classical (-like) error-prone variable). Other notations may be introduced when needed.

## 4. Development of Approximation Methods

### 4.1. In the Absence of Exposure Measurement Error

#### 4.1.1. Single Exposure Variable with Heteroskedastic Variance

By refining Lachin (2008)'s method, development begins with a single exposure variable $x_{t-l}$ as the only covariate. The standard error (*SE*) for the coefficient for $x_{t-l}$ in conditional logistic regression, $\theta_l$, which is used to estimate the effect of an exposure on an outcome is,

$$SE(\hat{\theta}_l) = \sqrt{I(\theta_l)^{-1}} \approx \sqrt{\sum_{h=1}^{n} \sigma_h^2(x_{t-l})}^{-1} \quad (\text{Eq.1})$$

where $I(\theta_l)$ denotes the Fisher Information (See Web Appendix 1 for the derivation). Thus, $\sigma_h^2(x_{t-l})$ is a key factor that determines the precision of the estimator and power. Suppose heteroskedastic $\sigma_h^2(x_{t-l})$ over $h$ because the variance of exposures can vary by space and time.

To derive the actual sample size, for computational convenience, let $\bar{\sigma}^2 := \frac{\sum_{h=1}^{n} \sigma_h^2(x_{t-l})}{n}$. Note that $\sigma_h^2(x_{t-l})$ is heteroskedastic across strata and $\bar{\sigma}^2$ is an average of heteroskedastic variances. In practice, $\bar{\sigma}^2$ may be calculated using $\sigma_h^2(x_{t-l})$ from external data about exposure distribution in a source population (See Web Appendix 2 for examples). The variance of each temporally lagged variable is almost identical to that of other: $V(x_{t-1}) \approx V(x_{t-l})$, so that $\sigma_h^2(x_t) \approx \sigma_h^2(x_{t-l})$. To provide a desired level of power $1 - \beta$ when using a test at level $\alpha$, $n$ is required as

$$n \cong \frac{(Z_{1-\alpha}+Z_{1-\beta})^2}{\theta_l^2 \bar{\sigma}^2} \quad (\text{Eq.2})$$

where $Z_{1-\alpha}$ is the critical value for the test (for a two-sided test, $Z_{1-\alpha/2}$). This is not a closed form because of $\bar{\sigma}^2$. An iterative process regarding $\sigma_h^2(x_{t-l})$ over strata, thus, $n$ and then calculating $\bar{\sigma}^2$ may be used to identify $n$. Recall that $n$ also indicates the number of strata because one stratum has only one case and recall that $h$ denotes $h$-th case among $n$ cases/strata. If investigators find that $\bar{\sigma}^2$ reaches a certain value over $n$ with different values of $\sigma_h^2(x_{t-l})$, Eq.2 would be treated to have a closed solution. In matched case-control studies when the exposure variance is constant, it is known that statistical power increases with the number of controls and the power gain diminishes as more controls are used. Eq.2 considers the number of controls through $\sigma_h^2(x_{t-l})$ and reveals that when the exposure is heteroskedastic, the power gain may increase with the number of controls if this leads to an increase in $\sigma_h^2(x_{t-l})$, assuming that the controls come from the same source population as the cases.

### 4.1.2. Covariate Adjustment

With multiple covariates, a deflation factor may be useful (Lachin, 2008) as shown in exponential regression (Bernardo et al. , 2000) and as covariate adjustment would reduce the precision of the estimator in logistic regression in observational studies (Robinson & Jewell, 1991). Let $R^2_{x_{t-l}|z}$ be the coefficient of determination in the regression of $x_{t-l}$ on $z$. Then,

$$SE(\hat{\theta}_l|z) \approx \sqrt{I(\theta_l)\left(1 - R^2_{x_{t-l}|z}\right)}^{-1} \quad (Eq.3)$$

$$n \cong \frac{(Z_{1-\alpha}+Z_{1-\beta})^2}{\theta_l^2 \bar{\sigma}^2 \left(1 - R^2_{x_{t-l}|z}\right)} \quad (Eq.4)$$

### 4.1.3. Distributed Lag Models

When the effect of $x_{t-l}$ on $Y$ is of interest, the exposure at other times, $\boldsymbol{x}^{-l}$ can be seen as confounders when their effects exist (Kim & Lee, 2019). Eqs.3–4 may be refined by adding $\boldsymbol{x}^{-l}$ to $\boldsymbol{z}$. On the other hand, investigators may wish to identify the effect of cumulative exposure, from day 0 to day $L$. Then, the estimand for the effect of cumulative exposure is $\theta_0 x_t + \theta_1 x_{t-1} + \cdots \theta_L x_{t-L} = \sum_{l=0}^{L} \theta_l x_{t-l}$. For high $L$, constrained distributed lag variables may be used(Gasparrini et al., 2010, Mork et al., 2024). For sample size calculation, it may be convenient to consider a weighted average as $\bar{x}_t = \sum_{l=0}^{L} \frac{\theta_l}{\bar{\theta}} x_{t-l}$ where $\bar{\theta} = \sum_{l=0}^{L} \theta_l$. For critical time-window identifications, the following may be alternatively used (Langholz et al., 1999), $\bar{x}_{t-l_s} = \sum_{l=l_s}^{L} w_l x_{t-l}$ where $w_l$ is a weight and $l_s > 0$. The weight may be pre-specified or estimated. For illustration simplicity, I adhere to $\bar{x}_t$. Let $\bar{\sigma}^2_{\bar{x}_t} = \frac{\sum_{h=1}^{n} \sigma_h^2(\bar{x}_t)}{n}$ and $I(\bar{\theta}) =$

$$\sum_{h=1}^{n} \left[ \frac{\sum_{j=1}^{m_h} \bar{x}_{h,j}^2 \exp(\bar{\theta} \bar{x}_{h,j})}{\sum_{j=1}^{m_h} \exp(\bar{\theta} \bar{x}_{h,j})} - \left( \frac{\sum_{j=1}^{m_h} \bar{x}_{h,j} \exp(\bar{\theta} \bar{x}_{h,j})}{\sum_{j=1}^{m_h} \exp(\bar{\theta} \bar{x}_{h,j})} \right)^2 \right] \approx n\bar{\sigma}^2_{\bar{x}_t}, \text{ then}$$

$$n \approx \frac{(Z_{1-\alpha}+Z_{1-\beta})^2}{\bar{\theta}^2 \bar{\sigma}^2_{\bar{x}_t} \left(1 - R^2_{\bar{x}_t|z}\right)} \quad (Eq.5)$$

$$SE\left(\hat{\bar{\theta}}|z\right) \approx \sqrt{I(\bar{\theta})\left(1 - R^2_{\bar{x}_t|z}\right)}^{-1} \quad (Eq.6)$$

When the weight is unknown, investigators may use different pre-specified weights and examine how sample size and precision may vary over weights.

### 4.2. Non-Differential Additive Errors

Non-differential exposure error means that the error is independent of $Y$. Exposure variables may be provided at locations (e.g., measurement at environmental monitors), spatial grids, or a geographical delineation (e.g., city, census boundary). When $x_{g,t}$ is used in the place of $x_{i,t}$, the resulting measurement error is Berkson error as $x_{i,t} = x_{g,t} + u^B_{i,t}$ and $u^B_{i,t} \sim N(0, \varphi^2_B)$ where $u^B_{i,t}$ is independent and identically distributed with constant $\varphi^2_B$. As illustrated in Section 3, I use the subscript $i$ or $g$ hereafter because measurement error is considered. The Berkson error does not introduce bias in effect estimates in linear regression. In non-linear models such as logistic regression, Berkson error introduces negligible bias, particularly if the effect size is small(Keogh et al. , 2020). Suppose a distributed lag model with $x_g$ and $z$ as covariates are fit to estimate $\theta_l$. The actual sample size is

$$n \cong \frac{(Z_{1-\alpha}+Z_{1-\beta})^2}{\theta_l^2 \bar{\sigma}^2_{x_{g,t-l}} \left(1-R^2_{x_{g,t-l}|x_g^{-l},z}\right)} \text{ (Eq.7)}$$

$$SE(\hat{\theta}_l|x_g^{-l},z) \approx \sqrt{n\bar{\sigma}^2_{x_{g,t-l}} \left(1-R^2_{x_{g,t-l}|x_g^{-l},z}\right)}^{-1} \text{ (Eq.8)}$$

where $\bar{\sigma}^2_{x_{g,t-l}} = \frac{\sum_{h=1}^n \sigma_h^2(x_{g,t-l})}{n}$. These equations consider multicollinearity of lagged variables using $R^2_{x_{g,t-l}|x_g^{-l},z}$ so that Eq.7 refers to the actual sample size, not the effective sample size. The variability of $x_{g,t-l}$ is smaller than the variability of $x_{i,t-l}$ by $\varphi^2_B$, which explains the precision reduction. Covariate measurement errors in $x_g^{-l}$ would not introduce residual confounding if the error is non-differential Berkson because $u^B_{i,t-l^*}$ is not correlated with $x_{g,t-l}$. Regression-calibrated exposure variables, if correctly constructed, can be used because they are non-differential Berkson error variables (Boe et al. , 2023).

Other error may arise. Exposure measurement may be error-prone, which is not uncommon in biomedical and epidemiologic research. Suppose a linear measurement error model, $x_{g,t}^{ep} = \gamma_0 + \gamma_1 x_{g,t} + u_{g,t}^L$ and $u_{g,t}^L \sim N(0, \varphi_L^2)$ where $u_{g,t}^L$ is independent and identically distributed with constant $\varphi_L^2$. For illustration, suppose the Berkson error is absent (i.e., $g = i$). In linear regression, this error results in either underestimation or overestimation of effects by a factor of $\lambda_l = \frac{\gamma_1 V(x_{g,t-l}|z)}{\gamma_1^2 V(x_{g,t-l}|z) + V(u_{g,t-l}^L|z)} = \frac{\gamma_1 V(x_{g,t-l}|z)}{V(x_{g,t-l}^{ep}|z)}$ which is called bias factor (Innes et al. , 2021). And $\theta_l^L = \lambda_l \theta_l$ where $\theta_l^L$ denotes the quantity biased by the error, assuming no residual confounding by $x_g^{-l}$ due to the use of $x_g^{ep-l}$ (i.e., covariate measurement errors). This assumption will be lifted later. $V(\cdot|z)$ denotes the residual variance of a variable conditional on $z$. For temporally lagged variables, it is likely $V(x_{g,t}|z) \approx V(x_{g,t-l}|z)$. $\lambda \theta_l$ is a good approximation in logistic regression if the measurement error is small, the effect size is small (Carroll et al. , 2006), or logistic regression in case-control analyses may be equal or similar to log-linear regression, depending on study settings (Lu & Zeger, 2007). How *small* the effect size or the measurement error is in relation to the variance of the exposure. When $\gamma_1 = 1$, the resulting error is classical error, which results in underestimation: $0 < \lambda_l = \frac{V(x_{g,t-l}|z)}{V(x_{g,t-l}^{ep}|z)} < 1$. $n$ approximates

$$n \cong \frac{(Z_{1-\alpha} + Z_{1-\beta})^2}{\lambda_l^2 \theta_l^2 \bar{\sigma}_{x_{g,t-l}^{ep}}^2 \left(1 - R_{x_{g,t-l}^{ep}|x_g^{ep-l},z}^2\right)} \quad (Eq.9)$$

$$SE\left(\hat{\theta}_l^L | x_g^{ep-l}, z\right) \approx \sqrt{n \bar{\sigma}_{x_{g,t-l}^{ep}}^2 \left(1 - R_{x_{g,t-l}^{ep}|x_g^{ep-l},z}^2\right)}^{-1} \quad (Eq.10)$$

Eq. 9 shows that the actual sample size, compared to no measurement error (i.e., $g = i$), is increased by the factor $\dfrac{\bar{\sigma}^2_{x_{g,t-l}}\left(1-R^2_{x_{g,t-l}|x_g^{-l},z}\right)}{\lambda_l^2 \bar{\sigma}^2_{x^{ep}_{g,t-l}}\left(1-R^2_{x^{ep}_{g,t-l}|x_g^{ep-l},z}\right)}$. If $\dfrac{\bar{\sigma}^2_{x_{g,t-l}}\left(1-R^2_{x_{g,t-l}|x_g^{-l},z}\right)}{\bar{\sigma}^2_{x^{ep}_{g,t-l}}\left(1-R^2_{x^{ep}_{g,t-l}|x_g^{ep-l},z}\right)}$ is assumed to be

approximately $\lambda_l = \dfrac{V(x_{g,t-l}|z)}{V(x^{ep}_{g,t-l}|z)}$, then, the sample size is approximately increased by $\dfrac{1}{\lambda_l} =$

$\dfrac{V(x^{ep}_{g,t-l}|z)}{V(x_{g,t-l}|z)} = \dfrac{1}{\rho^2_{x^{ep}_{g,t-l},x_{g,t-l}|z}}$, where $\rho^2_{x^{ep}_{g,t-l},x_{g,t-l}|z}$ is the squared partial correlation between $x_{g,t-l}$ and

$x^{ep}_{g,t-l}$. This agrees with the previous work (Keogh et al., 2020). If the error is linear, then, the

factor may be assumed to be $\dfrac{V(x_{g,t-l}|z)}{V(x^{ep}_{g,t-l}|z)} = \dfrac{\lambda_l}{\gamma_1}$. The sample size would be increased by $\dfrac{\gamma_1}{\lambda_l} =$

$\dfrac{V(x^{ep}_{g,t-l}|z)}{V(x_{g,t-l}|z)} = \dfrac{1}{\rho^2_{x^{ep}_{g,t-l},x_{g,t-l}|z}}$. Similarly, Eq.7 (Berkson error) is increased by the factor

$\dfrac{\bar{\sigma}^2_{x_{i,t-l}}\left(1-R^2_{x_{i,t-l}|z}\right)}{\bar{\sigma}^2_{x_{g,t-l}}\left(1-R^2_{x_{g,t-l}|x_g^{-l},z}\right)}$. Again, the factor may be approximately assumed to be $\dfrac{V(x_{i,t-l}|z)}{V(x_{g,t-l}|z)} =$

$\dfrac{1}{\rho^2_{x_{g,t-l},x_{i,t-l}|z}}$, where $\rho^2_{x_{g,t-l},x_{i,t-l}|z}$ is the squared partial correlation between $x_{i,t-l}$ and $x_{g,t-l}$. These

reveal that the squared partial correlation can be useful for calculating the sample size even when the variance is heteroskedastic. When both Berkson and linear (or classical) errors arise (i.e., $g \neq i$), the error becomes compound error (CE). Eqs. 9 and 10 subsume CE.

When residual confounding by $x_g^{-l}$ is not negligible, $\theta_l^L$ in Eq.9 needs to account for the residual confounding. This also implies that the squared partial correlation may not perform well. This is because $\theta_l^L$ is not simply proportional to $\lambda_l$. Web Appendix 3 develops an approximation equation for $\theta_l^L$ in the context of regression calibration (Boe et al., 2023). As will be shown, this may be the most accurate approximation equation in this paper, which is only available in practice when exposure validation data is available. Other approximation equations that can be used when exposure validation data is not available will be introduced later (Section 4.4). The squared partial correlation may not perform well when the error is not simply additive.

### 4.3. Complex Additive and Multiplicative Errors

#### 4.3.1. Autocorrelated Errors

Let $Q$ denote a set of causal factors of exposure of interest, which may exhibit autocorrelation (e.g., spatiotemporal patterns; within-individual or between individual patterns). For illustrational simplicity, we assume that within-individual variability (e.g., exposure variation over time) is subsumed under temporal autocorrelation, and between-individual variability is subsumed under spatial autocorrelation, as individuals may be in different locations.

For exposure prediction, autocorrelated error may stem from unmeasured or overlooked causal factors, or from $Q$ measured with error. For exposure measurement, autocorrelated error may arise from sources of error that could be correlated within or between individuals. Collectively, $Q^e$ denotes a set of sources of autocorrelation. For example, for air pollution estimates from prediction models, $Q^e$ may include imperfect documentations or measurements for emission activities, atmospheric process, solar radiation, and interaction with built-environment. $Q^e$ may also indicate behavioral factors such that outdoor air pollution levels may differ from personal exposure levels to these. For any cases where $g = i$ or $g \neq i$, $x_{g,t}^{ep}$ may be described as $x_{g,t}^{ep} = \gamma_0 + \sum_{j=0}^{J} \eta_j x_{g,t-j} + u_{g,t}^{LL}$. $\sum_{j=0}^{J} \eta_j x_{g,t-j}$ describes spatiotemporal autocorrelation process of $x_{g,t}^{ep}$ to some extent, analogous to autoregression. The notation, $x_{g,t}^{ep}$, is used to denote an exposure variable having at least non-Berkson error. When $g \neq i$, $x_{g,t}^{ep}$ denotes an exposure variable having both Berkson error and non-Berkson error. When $g = i$, $x_{g,t}^{ep}$ denotes an exposure variable having non-Berkson error. $u_{g,t}^{LL}$ may be autocorrelated and not be independent of $x_{g,t}$. Let $u_{g,t}^{LL} = ð(\boldsymbol{Q}^e) + u_{g,t}^{NI} + u_{g,t}^L$; $u_{g,t}^{NI} \sim \mathcal{G}$. $\mathcal{E}(\boldsymbol{Q}^e)$ indicates association between $\boldsymbol{Q}^e$ and $x_{g,t}^{ep}$ that explains spatiotemporal autocorrelation process to some extent. $u_{g,t}^{NI}$ is an error term indicating the remaining spatiotemporal autocorrelation, which follows an unknown multivariate distribution $\mathcal{G}$. The superscript, *NI*, indicates non-independence. Let this error model be referred to as a linear-like (LL) error model because this error model subsumes linear error (and classical error). For illustrational simplicity, suppose $x_{g,t}^{ep} = \gamma_0 + \gamma_1 x_{g,t} + u_{g,t}^{LL}$ by letting $u_{g,t}^{LL} \equiv \sum_{j=1}^{J} \eta_j x_{g,t-j} + ð(\boldsymbol{Q}^e) + u_{g,t}^{NI} + u_{g,t}^L$ and $\gamma_1 \equiv \eta_0$. The approximation equation developed in Web Appendix 3 and other approximations that will be introduced in Section 4.4 can be applied to $\theta_l^{LL}$, the quantity biased by LL error. Finally,

$$n \cong \frac{(Z_{1-\alpha}+Z_{1-\beta})^2}{\theta_l^{LL^2}\bar{\sigma}_{x_{g,t-l}^{ep}}^2\left(1-R^2_{x_{g,t-l}^{ep}|x_g^{ep^{-l}},z}\right)} \quad \text{(Eq.11)}$$

$$SE\left(\hat{\theta}_l^{LL}|\boldsymbol{x}_g^{ep^{-l}},\boldsymbol{z}\right) \approx \sqrt{n\bar{\sigma}_{x_{g,t-l}^{ep}}^2\left(1-R^2_{x_{g,t-l}^{ep}|x_g^{ep^{-l}},z}\right)^{-1}} \quad \text{(Eq.12)}$$

And for the estimator for cumulative exposure,

$$n = \frac{(Z_{1-\alpha}+Z_{1-\beta})^2}{\bar{\theta}^{LL^2}\bar{\sigma}_{\bar{x}_{g,t}^{ep}}^2\left(1-R^2_{\bar{x}_{g,t}^{ep}|z}\right)} \quad \text{(Eq.13)}$$

$$SE\left(\hat{\bar{\theta}}^{LL}|\boldsymbol{z}\right) \approx \sqrt{n\bar{\sigma}_{\bar{x}_{g,t}^{ep}}^2\left(1-R^2_{\bar{x}_{g,t}^{ep}|z}\right)^{-1}} \quad \text{(Eq.14)}$$

where $\bar{x}_{g,t}^{ep} = \sum_{l=0}^{L} \frac{\theta_l^{LL}}{\bar{\theta}^{LL}} x_{g,t-l}^{ep} \approx \sum_{l=0}^{L} \frac{\theta_l}{\bar{\theta}} x_{g,t-l}^{ep}$.

Berkson-like (BL) error may arise when $g \neq i$. For illustration, suppose the error model, $x_{i,t} = x_{g,t} + u_{i,t}^{BL}$ where $u_{i,t}^{BL} = u_{i,t}^B + u_{i,t}^{NI}$ and $u_{i,t}^{NI}$ denotes *NI* error. Unlike Berkson error, BL error may introduce bias in effect estimates because $u_{i,t}^{BL}$ may be correlated with $x_{g,t}$. Let $\theta_l^{BL}$ denote the quantity biased by the error, $\theta_l^{BL} \cong \theta_l + \sum_{k=0}^{L} \theta_k \varrho_{u_{i,t-k}^{BL}, x_{g,t-l} | x_g^{-l}, z}$, where

$\varrho_{u_{i,t-k}^{BL}, x_{g,t-l} | x_g^{-l}, z}$ is the regression coefficient for $x_{g,t-l}$ when $u_{i,t-k}^{BL}$ is regressed on $x_{g,t}$ and $z$. For $n$, Eqs.7–8 can be used by replacing $\theta_l$ with $\theta_l^{BL}$. When both BL and LL errors exist, Eqs.11–14 can be used by replacing $\theta_l^{LL}$ and $\bar{\theta}^{LL}$ with $\theta_l^{CE} \cong \theta_l^{LL} + \sum_{k=0}^{L} \theta_k \varrho_{u_{i,t-k}^{BL}, x_{g,t-l} | x_g^{-l}, z}$ and $\bar{\theta}^{CE} = \sum_{l=0}^{L} \theta_l^{CE}$. For exposure measurement error corrections, the bias due to BL error implies that a calibrated exposure variable should be constructed such that *NI* error becomes independent error. For example, regression calibration models may be constructed using spatiotemporal regression techniques.

### 4.3.2. Differential Errors

Differential exposure measurement error means that the error is not independent of $Y$. There are at least five mechanisms through which the error can become differential: 1) the correlation between $x_{i \text{ or } g,t}$ and $u_{i \text{ or } g,t}^{LL}$ through which $u_{i \text{ or } g,t}^{LL}$ is correlated with $Y$ when $x_{i \text{ or } g,t}$ has effects on $Y$ but $x_{i \text{ or } g,t}^{ep}$ is used to estimate the effect of exposure (Figure 1A); 2) the correlation between $x_{g,t}$ and $u_{i,t}^{BL}$ through which $u_{i,t}^{BL}$ is correlated with $Y$ when $x_{i,t}$ has effects on $Y$ but $x_{g,t}$ is used to estimate the effect of exposure (Figure 1B); 3) $u_{i \text{ or } g,t}^{LL}$ or $u_{g,t}^{BL}$ as a risk factor of $Y$ or related to non-confounder risk factors of $Y$ (Figures 1C–D); 4) the correlation of $u_{i \text{ or } g,t}^{LL}$, or $u_{g,t}^{BL}$ with unmeasured confounders; and 5) the correlation of $u_{i \text{ or } g,t}^{LL}$, or $u_{g,t}^{BL}$ with confounder measurement error when a confounder is measured with error. Thus, BL and LL errors would be differential if there exist exposure effects.

Potential bias to the estimators needs to be considered to accurately calculate sample size. Differential errors in distribute lag models may mitigate residual confounding due to the use of $x_g^{ep-l}$ (See Web Appendix 4 for details). Differential errors may be (coincidently) adjusted for by using matching or adjustment for covariates that makes the error differential, so that the resulting error may become conditionally non-differential. Variables or statistical terms for adjustment for differential errors could be subsumed into $\mathbf{z}$ (See Web Appendix 5 for details).

### 4.3.3. Multiplicative Errors

Multiplicative errors may be transformed to additive errors. For example, multiplicative Berkson (-like) (MB(L)) error may be described as $\log(x_{i,t}) = \log(x_{g,t}) + u_{i,t}^{B(L)}$ and $x_{i,t} = x_{g,t}\exp(u_{i,t}^{B(L)})$. If $\log(x_{g,t})$ is used as the exposure variable in an outcome model, the consequence would be identical to that for the additive error, assuming that $\log(x_{g,t})$ can correctly represent $x - Y$ relationship. If $x_{g,t}$ is used and the error is not small, the consequence may differ. If residual confounding due to the use of $\mathbf{x}_g$ is negligible, $\theta_l^{MB(L)}$, the quantity biased by the error would be $\theta_l^{MB(L)} \approx \theta_l E\left[\exp\left(u_{i,t}^{B(L)}\right)\right]$. Note $E\left[\exp\left(u_{i,t}^{B(L)}\right)\right] \neq \exp\left(E\left[u_{i,t}^{B(L)}\right]\right) = 1$.

Multiplicative linear (-like) (L(L)) error may be described as $\log(x_{g,t}^{ep}) = \gamma_{m0} + \gamma_{m1}\log(x_{g,t}) + u_{g,t}^{L(L)}$ and $x_{g,t}^{ep} = (x_{g,t})^{\gamma_{m1}}\exp(\gamma_{m0})\exp(u_{g,t}^{L(L)})$. If $x_{g,t}^{ep}$ is used as the exposure variable and the error is not small, the consequence would differ. For example, for multiplicative classical (MC) error (i.e., $\gamma_{m1} = 1$ and $\gamma_{m0} = 0$), assuming that the residual confounding due to the use of $\mathbf{x}_g^{ep-l}$ is negligible, $\theta_l^{MC} \cong \theta_l \dfrac{V(x_{g,t-l}|\mathbf{z})}{V(x_{g,t-l}|\mathbf{z})\left(1+V(u_{g,t}^L|\mathbf{z})\right)+E[x_{g,t-l}|\mathbf{z}]^2 V(u_{g,t}^L|\mathbf{z})}$

demonstrating attenuation. To approximate $\theta_l^{ML(L)}$, the quantity biased by the multiplicative L(L) (ML(L)) error, the approximation equation in Web Appendix 3 can be used for different error models (Web Appendix 5) but requires the use of validation data. Approximation equations that will be introduced in Section 4.4, which does not require exposure validation data, may also be used.

Approximation of the variance of the estimator can also be obtained. The standard error of $\theta_l^{ML(L)}$ may be approximated as

$$SE\left(\hat{\theta}_l^{ML(L)}|z, x_g^{ep-l}\right) \approx$$

$$\sqrt{\left(n\bar{\sigma}^2_{x^{ML(L)}_{g,t-l}}\left(1 - R^2_{x^{ep}_{g,t-l}|x^{ep-l}_g,z}\right)\right)^{-1} + \theta_l^{ML(L)^2}\left(n\bar{\sigma}^2_{\exp(\gamma_{m0})\exp(u^{L(L)}_{g,t})}\left(1 - R^2_{x^{ep}_{g,t-l}|x^{ep-l}_g,z}\right)\right)^{-1}} \quad (\text{Eq.15})$$

where let $x_{g,t-l}^{ML(L)} = (x_{g,t-l})^{\gamma_{m1}}$ (See Web Appendix 6 for the derivation). The term $(x_{g,t-l})^{\gamma_{m1}}$ demonstrates the standard error can be either attenuated or inflated, depending on $\gamma_{m1}$. The inflation may arise if $\gamma_{m1} < 1$ when $x_{g,t-l} > 1$. The second component in the square root demonstrates that the standard error may be inflated, but this component could be relatively small compared to the first component, especially for small effect size, so that this may be negligible. Similarly, the standard error of $\bar{\theta}^{ML(L)}$, which is the quantity regarding the cumulative exposure effect, may be approximated as

$$SE\left(\hat{\bar{\theta}}^{ML(L)}|z\right) \cong \sqrt{\left(n\bar{\sigma}^2_{\bar{x}^{ML(L)}_{g,t}}\left(1 - R^2_{\bar{x}^{ep}_{g,t}|z}\right)\right)^{-1} + \bar{\theta}^{ML(L)^2}\left(n\bar{\sigma}^2_{\exp(\gamma_{m0})\exp(\bar{u}^{L(L)}_{g,t})}\left(1 - R^2_{\bar{x}^{ep}_{g,t}|z}\right)\right)^{-1}} \quad (\text{Eq.16})$$

where $\bar{x}_{g,t}^{ML(L)} = \exp(\gamma_{m0})E\left[\exp\left(u_{g,t}^{L(L)}\right)\right]\left(\frac{\theta_0}{\bar{\theta}}(x_{g,t})^{\gamma_{m1}} + \cdots + \frac{\theta_L}{\bar{\theta}}(x_{g,t-L})^{\gamma_{m1}}\right)$ and $\bar{u}_{g,t}^{L(L)} = \frac{\theta_0}{\bar{\theta}}u_{g,t}^{L(L)} + \cdots + \frac{\theta_L}{\bar{\theta}}u_{g,t}^{L(L)}$. $n$ approximates

$$n \cong \frac{(Z_{1-\alpha} + Z_{1-\beta})^2}{\theta_l^{ML(L)^2}\bar{\sigma}^2_{x^{ML(L)}_{g,t-l}}\left(1 - R^2_{x^{ep}_{g,t-l}|x^{ep-l}_g,z}\right)VCF} \quad (\text{Eq.17})$$

$$n \cong \frac{(Z_{1-\alpha} + Z_{1-\beta})^2}{\bar{\theta}^{ML(L)^2}\bar{\sigma}^2_{\bar{x}^{ML(L)}_{g,t}}\left(1 - R^2_{\bar{x}^{ep}_{g,t}|z}\right)VCF} \quad (\text{Eq.18})$$

where $VCF$ denotes a variance correction factor,

$$VCF = \frac{\left(n\bar{\sigma}^2_{x^{ML(L)}_{g,t-l}}\right)^{-1}}{\left(n\bar{\sigma}^2_{x^{ML(L)}_{g,t-l}}\right)^{-1} + \theta_l^{ML(L)^2}\left(n\bar{\sigma}^2_{\exp(\gamma_{m0})\exp(u^{L(L)}_{g,t})}\right)^{-1}} \text{ or}$$

$$VCF = \frac{\left(n\bar{\sigma}^2_{\bar{x}^{ML(L)}_{g,t}}\right)^{-1}}{\left(n\bar{\sigma}^2_{\bar{x}^{ML(L)}_{g,t}}\right)^{-1} + \bar{\theta}^{ML(L)^2}\left(n\bar{\sigma}^2_{\exp(\gamma_{m0})\exp(\bar{u}^{L(L)}_{g,t})}\right)^{-1}}$$

for small $\theta_l^{ML(L)}$ or $\bar{\theta}^{ML(L)}$, $VCF \approx 1$. For multiplicative CE (i.e., $g \neq i$), Eqs.15–18 can be used by considering additional bias due to BL error to the estimator. For only MBL error, Eqs. 7–10 can be used.

### 4.4. When Exposure Validation Data is Not Available

#### 4.4.1. $\bar{\sigma}^2$ Calculation

For additive errors, the exposure variable measured/estimated with error that investigators have, which would be $x_{g,t}$ for B(L) error or $x^{ep}_{g,t}$ for non-Berkson error or CE, can be directly used to calculate $\bar{\sigma}^2$ (Note Eqs.7–14 that are based on $x_{g,t}$ or $x^{ep}_{g,t}$). For ML(L), what investigators have would be $x^{ep}_{g,t}$. $x^{ep}_{g,t}$ needs to be divided by the square root of $\exp(\gamma_{m0})\exp\left(u^{L(L)}_{g,t}\right)$, which is unknown without validation data. Since the variance decomposition for $(x_{g,t})^{\gamma_{m1}}\exp(\gamma_{m0})\exp\left(u^{L(L)}_{g,t}\right)$ may vary by the distribution of $x_{g,t}$ and $u^{L(L)}_{i,t}$, the generalizable decomposition may not be obtainable if $u^{L(L)}_{i,t}$ is unknown. Alternatively, it may be reasonable to consider some factor $c$ as

$$\bar{\sigma}^2_{x^{ML(L)}_{g,t-l}} \cong \bar{\sigma}^2_{x^{ep}_{g,t-l}} c$$

and conduct numerical analyses about how sample size numbers could vary by several $c$ values. Or investigators may, sometimes, have information about the error from external sources such as published documents, which may be used. Suppose that the coefficient of determination in the regression of the logarithm of $x_{g,t}^{ep}$ against the logarithm of $x_{g,t}$ ($R^2_{\log(x_{g,t}^{ep}),\log(x_{g,t})}$) and $\gamma_{m1}^{crude}$ as the regression coefficient are known and the coefficient in the regression of $x_{g,t}$ against $x_{g,t}^{ep}$ are known. Recall the following ML(L) model, $\log(x_{g,t}^{ep}) = \gamma_{m0} + \gamma_{m1} \log(x_{g,t}) + u_{g,t}^{L(L)}$ and $\log(x_{g,t-l}^{ep}) = \gamma_{m0} + \gamma_{m1} \log(x_{g,t-l}) + u_{g,t-l}^{L(L)}$ and,

$$R^2_{\log(x_{g,t}^{ep}),\log(x_{g,t})} \cong R^2_{\log(x_{g,t-l}^{ep}),\log(x_{g,t-l})} = \frac{V(\gamma_{m0} + \gamma_{m1}\log(x_{g,t-l}))}{V(\log(x_{g,t-l}^{ep}))} = \frac{V(\gamma_{m1}\log(x_{g,t-l}))}{V(\log(x_{g,t-l}^{ep}))}$$

for $\gamma_{m1} > 0$, $\bar{\sigma}^2_{x_{g,t-l}^{ML(L)}} \approx \bar{\sigma}^2_{\log(x_{g,t-l}^{ep})} E[x_{g,t-l}]^{2\gamma_{m1}} R^2_{\log(x_{g,t-l}^{ep}),\log(x_{g,t-l})}$

(Web Appendix 7 for the derivation). Thus, by assuming $\gamma_{m1}^{crude} \cong \gamma_{m1}$, which may be reasonable in practice when investigators have domain knowledge about exposure measurements or estimations, and calculating $E[x_{g,t-l}]$ based on the information of the regression of $x_{g,t}$ against $x_{g,t}^{ep}$, they can calculate $\bar{\sigma}^2_{\log(x_{g,t-l}^{ep})}$ with their data and then approximate $\bar{\sigma}^2_{x_{g,t-l}^{ML(L)}}$. $\bar{\sigma}^2_{\bar{x}_{g}^{ML(L)}}$ can also be approximated similarly. Sometimes, $R^2_{\log(x_{g,t}^{ep}),\log(x_{g,t})}$ is not known, but $R^2_{x_{g,t-l},x_{g,t-l}^{ep}}$, which may be used instead.

### 4.4.2. Bias Consideration

For non-Berkson error, the estimator may be biased so that investigators should consider the size of bias when calculating $n$. Recall the additive L(L) error model, $x_{g,t}^{ep} = \gamma_0 + \gamma_1 x_{g,t} + u_{g,t}^{LL}$ If $\gamma_1^{crude}$ and $R^2_{x_{g,t}^{ep},x_{g,t}}$ are known, by assuming $\gamma_1^{crude} = \gamma_1$, in distributed lag models,

$$\theta_l^{L(L)} \approx R^2_{x_{g,t}^{ep},x_{g,t}} \left( \frac{\theta_l \pm \left(1 - R^2_{x_{g,t}^{ep},x_{g,t}}\right) \sum_{j \neq l}^{L} \theta_j}{\gamma_1} \right) \quad (\text{Eq.19})$$

and this approximation may perform reasonably if residual confounding due to the use of $x_g^{ep-l}$ is not severe (Web Appendix 8 for the derivation). If "+" is used for "±" in the numerator, Eq.19 may indicate the upper bound of $\theta_l^{L(L)}$ to identify whether $\theta_l^{L(L)} \leq \theta_l$ is plausible but may result in underapproximation of sample size. If $x_{g,t}$ is highly autocorrelated and $x_{g,t}^{ep}$ is a good measure of $x_{g,t}$, "-" may be more accurate than "+" (Web Appendix 8). If residual confounding is negligible, or $\bar{\theta}$ is of interest,

$$\theta_l^{L(L)} \approx \frac{R^2_{x_{g,t}^{ep}, x_{g,t}} \theta_l}{\gamma_1} \text{ or } \bar{\theta}^{L(L)} \approx \frac{R^2_{x_{g,t}^{ep}, x_{g,t}} \bar{\theta}}{\gamma_1} \quad (Eq.20)$$

Multiplicative error may be treated as additive error. The ML(L) error model may be re-written as $x_{g,t}^{ep} = \gamma_0^{ML(L)} + \gamma_1^{crude} x_{g,t} + u_{g,t}^{ML(L)}$ where $u_{g,t}^{ML(L)}$ mimicks $u_{g,t}^{L(L)}$ in which an interaction between $x_{g,t}$ and another error component(s) exists but is unknown. Validation studies may report $\gamma_1^{crude}$ and $R^2_{x_t^{ep}, x_t}$ when whether the error is additive or multiplicative remains uncertain. Eq.20 relies on the assumption $R^2_{x_{g,t}^{ep}, x_g} \approx R^2_{x_{g,t-l}^{ep}, x_{g,t-l} | x_g^{ep-l}, z}$. It is not unrealistic that $R^2_{x_{g,t}^{ep}, x_g} \gg R^2_{x_{g,t-l}^{ep}, x_{g,t-l} | x_g^{ep-l}, z}$ in practice (Web Appendix 8). If $R^2_{x_{g,t-l}^{ep}, x_{g,t-l} | x_g^{ep-l}, z}$ is not known, it may be wise to apply a multiplication factor to $R^2_{x_{g,t}^{ep}, x_g}$ and examine how sample size can vary accordingly by changing that factor.

### 4.5. Polynomials for Non-Linear Effects

See Web Appendix 9.

### 5. Simulation Experiments

See Web Appendix 10 for details. Findings demonstrate that the calculations perform well. Eqs.19 and 20 may sometimes provide accurate approximates for effect estimates, suggesting that it would be difficult to accurately consider bias in the estimator without validation data when residual confounding due to distributed lags measured with error is not small.

## 6. Application

Suppose that investigators want to design a prospective or retrospective birth cohort study based on seven counties in the Greater Chicago Region to test whether prenatal air pollution exposures increase the risk of a health outcome of infants. Following the gestational age, investigators may link weekly air pollution levels to a cohort data based on gestational weeks and plan to conduct a matched nested case-control analysis. To calculate sample size, assume that the effect size for a 1-unit increase in air pollution levels during a gestational week (e.g., $1\mu g/m^3$ in $PM_{2.5}$) is 0.01 (i.e., $\theta_l = 0.01$). The deflation factor (i.e., $R^2_{x^{ep}_{g,t-l}|x^{ep-l}_g, z}$) for covariate adjustment may be assumed to range from 0.25 to 0.75, which may vary by a matching factor and the degree of other confounding adjustment including distributed lags (i.e., air pollution levels during other gestational weeks). After calculating weekly $PM_{2.5}$ levels using the United States Environmental Protection Agency's FAQSD modelled estimates (Reff, 2023) for the years 2015–2018, $\bar{\sigma}^2$ was calculated by picking one census tract randomly for a potential case and picking another census tract randomly for a potential control matched (one-to-one matching) or other census tracts randomly for potential controls matched (one-to-$k$ matching) several times, since exact spatiotemporal exposure distribution is unknown before looking at data or recruiting individuals. Thus, a distribution of $\bar{\sigma}^2$ as the average of heteroskedastic variance across strata was obtained. We may obtain a range of sample size required to achieve 80% statistical power by assuming that 1) no bias from exposure measurement error (e.g., no error or Berkson error); 2) bias from linear (-like) error based on Figure S1H (~4% overestimation); 3) additional bias from residual confounding due to linear (-like) error in air pollution levels during two other gestational weeks, and 4) additional bias from residual confounding due to linear (-like) error in air pollution levels during four other gestational weeks. Comparing our method with the assumption #1 (Figure 2A) to the traditional method by Lachin 2008 (Figure 2D) demonstrates that heteroskedasticity matters. In this application example, the traditional method overestimates the number of cases needed when the number of controls matched is ≤2 and either underestimate or overestimate the number of cases needed when the number of controls exceeds 2, also leading to an overestimation of the number of controls required. And under heteroskedasticity, additional controls do not impact sample size meaningfully. Figures 2B and 2C illustrate how *n* calculations can vary depending on the consideration of bias due to exposure measurement error. Figure 2B represents calculations without bias from residual confounding caused by linear(-like) errors in other covariates, while Figure 2C incorporates this bias. The comparison highlights that although the required sample size may decrease slightly due to an approximately 4% overestimation of effect estimates (Figure 2B)—a factor some may dismiss as irrelevant, considering it an artificial gain from overestimation—the presence of residual confounding may cause substantial toward-the-null-bias in effect estimates. Consequently, the required sample size–the number of cases and controls–may become much larger.

Figure 3 demonstrates how *n* calculation is sensitive to the degree of exposure measurement error, such as $R^2_{x_t^{ep}, x_t}$ and $\gamma_1$, with the assumptions #3 (Figure 3A) and #4 (Figure 3B) when $\theta_l$ is underestimated due to errors (i.e., $\theta_l^{L(L)}$) and $R^2_{x_{g,t-l}^{ep} | x_g^{ep-l}, z} = 0.5$. Again, the required *n* may become larger with the greater degree of residual confounding. Figures 3C and 3D is a version of Figures 3A and 3B when $\theta_l = 0.05$, not 0.01, implying scenarios that the actual effect size is five times. Approximately 1/50th of *n* may be needed. These may alternatively be interpreted as separate scenarios (#5) that the substantial bias due to exposure measurement error (including residual confounding) made $\theta_l^{L(L)} = 0.01$ but the bias was corrected using regression calibration while all other things are approximately equal.

In Figure 3, there exists the number of cases when $R^2_{x_t^{ep},x_t} = 1$ and $\gamma_1 = 1$ meaning no measurement error at all. So, readers may compare *n* in this scenario (i.e., the exposure variable is not measured with error) to other scenarios when the exposure variable is measured with error.

Investigators may be able to design a validation study to correct exposure measurement errors (e.g., regression calibration). In this case, assumption 1) with different values of $R^2_{x_t^{ep},x_t}$ may alternatively be considered as a calibration scenario because a regression calibrated (rc) exposure variable would be seen as the variable having Berkson error, $x_{i,t-l}^{rc}$. Investigators may wonder how much sample size gain is expected. In this application example, even if the residual confounding is absent (Figures 2A and 2B), the gain may become larger if the value of $R^2_{x_{i,t-l}^{rc},x_{i,t-l}}$ increases due to regression calibration. This may occur if **z** is correlated with *x* because **z** should be included into the calibration model, meaning that the correlation between $x_{i,t-l}^{rc}$ and $x_{i,t-l}$ may become higher than the correlation between $x_{g,t-l}$ and $x_{i,t-l}$. If the residual confounding is present, the gain may be substantial (Figures 2A, 2C). If the error correction leads to bias reduction substantially, say, $R^2_{x_{i,t-l}^{rc},x_{i,t-l}}=0.9$, the gain may also be substantial (Figure 3B). On top of this, if the effect estimate is expected to be 0.05, the gain may be very substantial (the left green box in Figure 3C). Thus, in certain situations, designing a validation study can be more cost-effective than collecting additional samples for the main study.

## 7. Discussion

A deeper understanding of the impact of measurement errors on sample size and the accuracy and precision of estimators is essential for designing robust, reproduceable research and for interpreting findings. Also, in disciplines where collecting large sample is challenging and/or the effect size to be tested is not large, it is critical to accurately calculate the required sample size. Several key lessons are summarized here. First, if exposure levels vary spatiotemporally, the variance may be heteroskedastic. This needs to be appropriately considered in understanding sample size. Under heteroskedasticity, additional controls may not necessarily increase statistical power (i.e., may not reduce the required number of cases), which contrasts with the fact that additional controls do increase statistical power when the exposure is binary or homoscedastic. Second, even without exposure validation data, investigators may identify sample size for a desired power and bias to the estimators based on external information about exposure measurement error. Third, for distributed lag models, it may sometimes be difficult to consider the accuracy of the estimators of lag variables due to potentially severe residual confounding from covariate measurement errors. For example, theoretically, the absence of associations between certain lag times and *Y* may be attributed to residual confounding. In identifying critical exposure time-windows, designing validation research may be desirable, but sample size could be still efficiently calculated. Fourth, in situations where validation data is unavailable, conducting numerical analyses to examine how sample size and the accuracy of estimators can vary across different bias scenarios would be beneficial for designing research and interpreting findings. If exposure validation data is available, regression-calibrated exposure variables can be used in sample size calculations because they are variables with Berkson error. Under certain circumstances, conducting a validation study may be more cost-effective than increasing the sample size in the main study for hypothesis testing. The sample size and bias approximations developed in this paper may be used to efficiently design research in various settings, including unconditional logistic regression and linear regression, with minor modifications (e.g., the incidence of $Y$ to the denominator for logistic regression; $\sigma_Y^2$ to the numerator for linear regression with continuous $Y$; the variance of exposure variables for all individuals/groups). As shown in Section 6, inappropriate consideration of these factors can lead to over-calculation of sample size–the number of both cases and controls. This may potentially result in wasted resources, ethical concerns, and statistical challenges.

The present work could be extended in a number of directions. The methods could be directly used as forms of uncertainty analyses, such as quantitative bias analyses. Numerical analyses that account for the sampling variability of a validation study could be incorporated. Additionally, the treatment of measurement errors in distributed lags, as discussed in this paper, could be broadened to encompass covariate measurement errors more generally. Various measurement error correction methods could also be explored. Another promising avenue is the application of a deeper understanding of measurement errors to integrate evidence from individual studies (e.g., meta-analyses) for hypothesis testing and replication (Brakenhoff et al. , 2018, Loken & Gelman, 2017). Heterogeneity in findings, if not adequately understood, may impede causal inference and evidence synthesis.

**Acknowledgement**

This work was supported by Basic Science Research Program through the National Research Foundation of Korea (NRF) funded by the Ministry of Education (2021R1A6A3A14039711), by University of Illinois Chicago School of Public Health Seed Funding Program, and by the National Institute of Environmental Research (NIER) funded by the Ministry of Environment (MOE) of the Republic of Korea (NIER-2021-03-03-007)

**References**

BERNARDO, M. P., LIPSITZ, S. R., HARRINGTON, D. P. & CATALANO, P. J. (2000). Sample size calculations for failure time random variables in non-randomized studies. *Journal of the Royal Statistical Society Series D: The Statistician* **49**, 31-40.

BOE, L. A., SHAW, P. A., MIDTHUNE, D., GUSTAFSON, P., KIPNIS, V., PARK, E., SOTRES-ALVAREZ, D., FREEDMAN, L., OF THE STRATOS INITIATIVE, O. B. O. T. M. E. & GROUP, M. T. (2023). Issues in Implementing Regression Calibration Analyses. *American Journal of Epidemiology* **192**, 1406-1414.

BRAKENHOFF, T. B., MITROIU, M., KEOGH, R. H., MOONS, K. G., GROENWOLD, R. H. & VAN SMEDEN, M. (2018). Measurement error is often neglected in medical literature: a systematic review. *Journal of clinical epidemiology* **98**, 89-97.

CARROLL, R. J., RUPPERT, D., STEFANSKI, L. A. & CRAINICEANU, C. M. (2006). *Measurement error in nonlinear models: a modern perspective*: Chapman and Hall/CRC.

DANIELI, C., SHEPPARD, T., COSTELLO, R., DIXON, W. G. & ABRAHAMOWICZ, M. (2020). Modeling of cumulative effects of time-varying drug exposures on within-subject changes in a continuous outcome. *Statistical methods in medical research* **29**, 2554-2568.

FEWELL, Z., DAVEY SMITH, G. & STERNE, J. A. (2007). The impact of residual and unmeasured confounding in epidemiologic studies: a simulation study. *American journal of epidemiology* **166**, 646-655.

GANJAVI, M. & FARAJI, B. (2019). Late effect of the food consumption on colorectal cancer rate. *International journal of food sciences and nutrition* **70**, 98-106.

GASPARRINI, A., ARMSTRONG, B. & KENWARD, M. G. (2010). Distributed lag non-linear models. *Statistics in medicine* **29**, 2224-2234.

GREENLAND, S. & ROBINS, J. M. (1985). Confounding and misclassification. *American journal of epidemiology* **122**, 495-506.

INNES, G. K., BHONDOEKHAN, F., LAU, B., GROSS, A. L., NG, D. K. & ABRAHAM, A. G. (2021). The measurement error elephant in the room: challenges and solutions to measurement error in epidemiology. *Epidemiologic reviews* **43**, 94-105.

KAIER, K., FRANK, U., HAGIST, C., CONRAD, A. & MEYER, E. (2009). The impact of antimicrobial drug consumption and alcohol-based hand rub use on the emergence and spread of extended-spectrum β-lactamase-producing strains: a time-series analysis. *Journal of Antimicrobial Chemotherapy* **63**, 609-614.

KEOGH, R. H., SHAW, P. A., GUSTAFSON, P., CARROLL, R. J., DEFFNER, V., DODD, K. W., KÜCHENHOFF, H., TOOZE, J. A., WALLACE, M. P. & KIPNIS, V. (2020). STRATOS guidance document on measurement error and misclassification of variables in observational epidemiology: part 1—basic theory and simple methods of adjustment. *Statistics in medicine* **39**, 2197-2231.

KIM, H. & LEE, J.-T. (2019). On inferences about lag effects using lag models in air pollution time-series studies. *Environmental Research* **171**, 134-144.

LACHIN, J. M. (2008). Sample size evaluation for a multiply matched case–control study using the score test from a conditional logistic (discrete Cox PH) regression model. *Statistics in medicine* **27**, 2509-2523.

LANGHOLZ, B., THOMAS, D., XIANG, A. & STRAM, D. (1999). Latency analysis in epidemiologic studies of occupational exposures: application to the Colorado Plateau uranium miners cohort. *American journal of industrial medicine* **35**, 246-256.

LEE, J.-T. & SCHWARTZ, J. (1999). Reanalysis of the effects of air pollution on daily mortality in Seoul, Korea: A case-crossover design. *Environmental health perspectives* **107**, 633-636.

LOKEN, E. & GELMAN, A. (2017). Measurement error and the replication crisis. *Science* **355**, 584-585.

LU, Y. & ZEGER, S. L. (2007). On the equivalence of case-crossover and time series methods in environmental epidemiology. *Biostatistics* **8**, 337-344.

MACLURE, M. (1991). The case-crossover design: a method for studying transient effects on the risk of acute events. *American journal of epidemiology* **133**, 144-153.

MITTLEMAN, M. A. & MOSTOFSKY, E. (2014). Exchangeability in the case-crossover design. *International journal of epidemiology* **43**, 1645-1655.

MORK, D., KIOUMOURTZOGLOU, M.-A., WEISSKOPF, M., COULL, B. A. & WILSON, A. (2024). Heterogeneous distributed lag models to estimate personalized effects of maternal exposures to air pollution. *Journal of the American Statistical Association* **119**, 14-26.

ROBINSON, L. D. & JEWELL, N. P. (1991). Some surprising results about covariate adjustment in logistic regression models. *International Statistical Review/Revue Internationale de Statistique*, 227-240.

TOSTESON, T. D., BUZAS, J. S., DEMIDENKO, E. & KARAGAS, M. (2003). Power and sample size calculations for generalized regression models with covariate measurement error. *Statistics in Medicine* **22**, 1069-1082.

WHITE, E., KUSHIZ, L. H. & PEPE, M. S. (1994). The effect of exposure variance and exposure measurement error on study sample size: implications for the design of epidemiologic studies. *Journal of clinical epidemiology* **47**, 873-880.

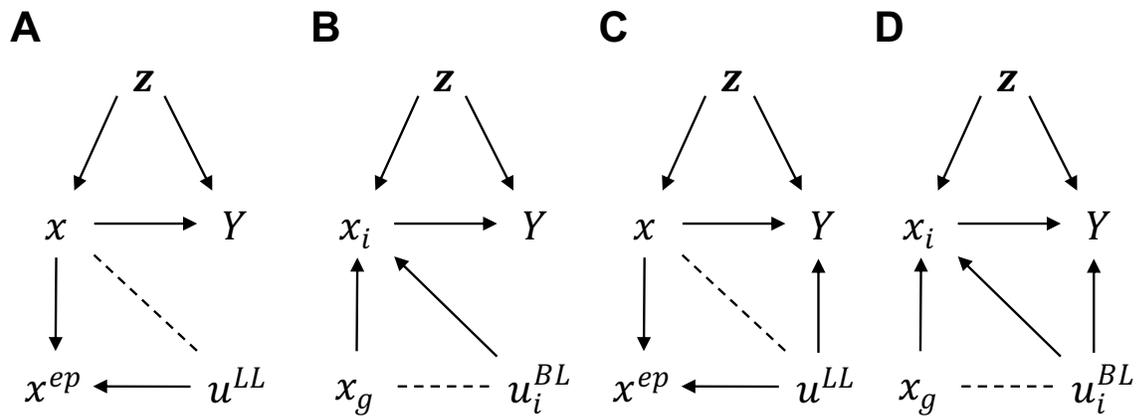

**Figure 1. Graphical descriptions of linear-like (LL) exposure measurement error (A, C) and Berkson-like (BL) exposure measurement error (B, D)**
**Note.** Dashed lines denote potential correlation.

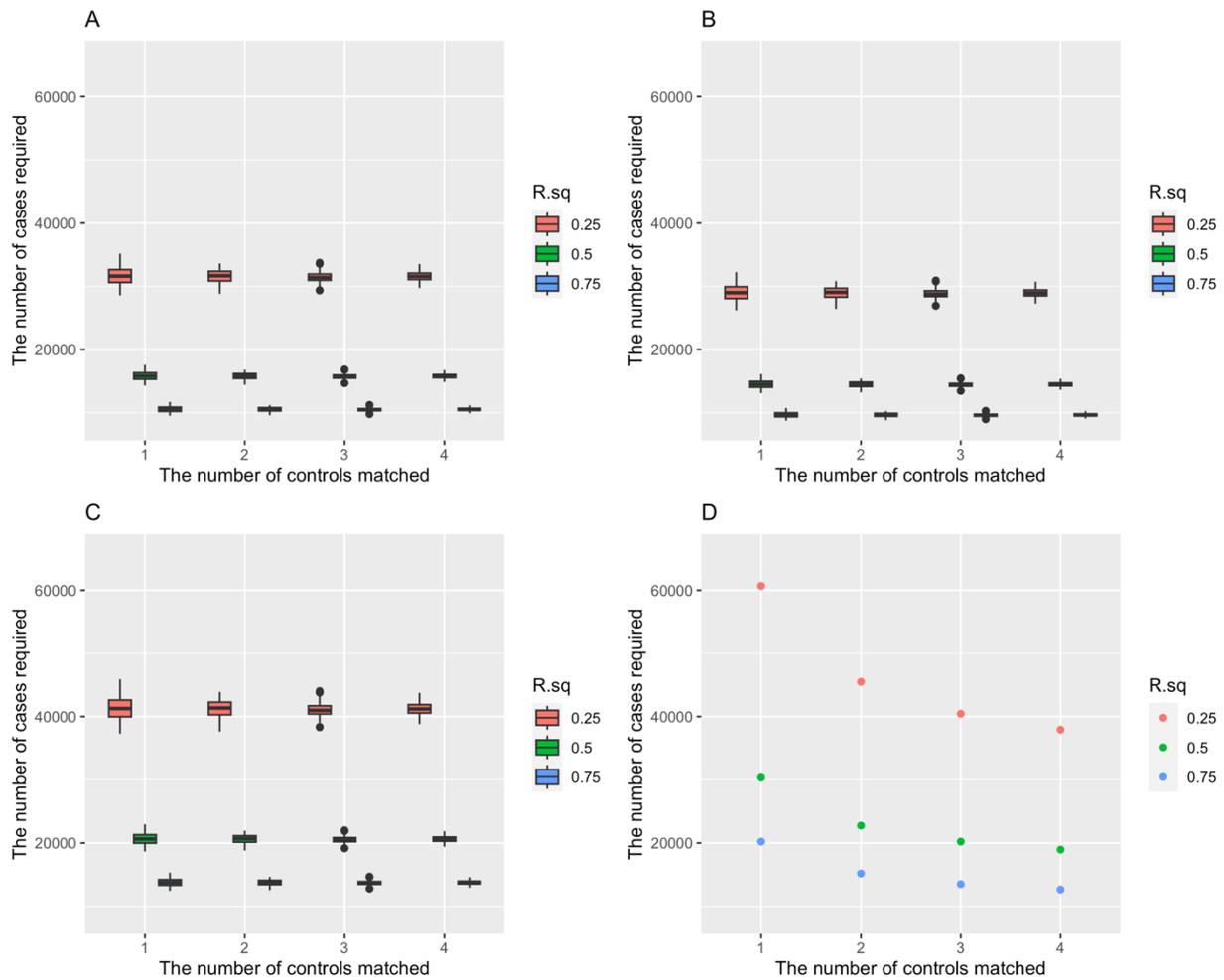

**Figure 2. *n* calculations based on the method developed in this study (A, B, C) and the traditional method by Lachin (2008) (D)**

Note. R.sq means $R^2_{x^{ep}_{g,t-l}|x^{ep-l}_g,z}$. Panel A shows sample sizes with the assumption that there is no bias in effect estimates (e.g., no error or Berkson error). Panel B shows sample sizes with the assumption that there is bias due to linear-like error, based on Figgure S1H (i.e., $R^2_{x^{ep}_{g,t},x_{i,t}} = 0.919$, and $\gamma_1 = 0.88$ which may result in overestimation by ~4%, where *g* may or may not equal *i*). Panel C shows sample sizes with the assumption that there is additional bias due to residual confounding due to linear-like error in exposures during two other gestational weeks. Panel D shows sample size calculations that do not account for exposure measurement errors, distributed lags, or heteroskedasticity.

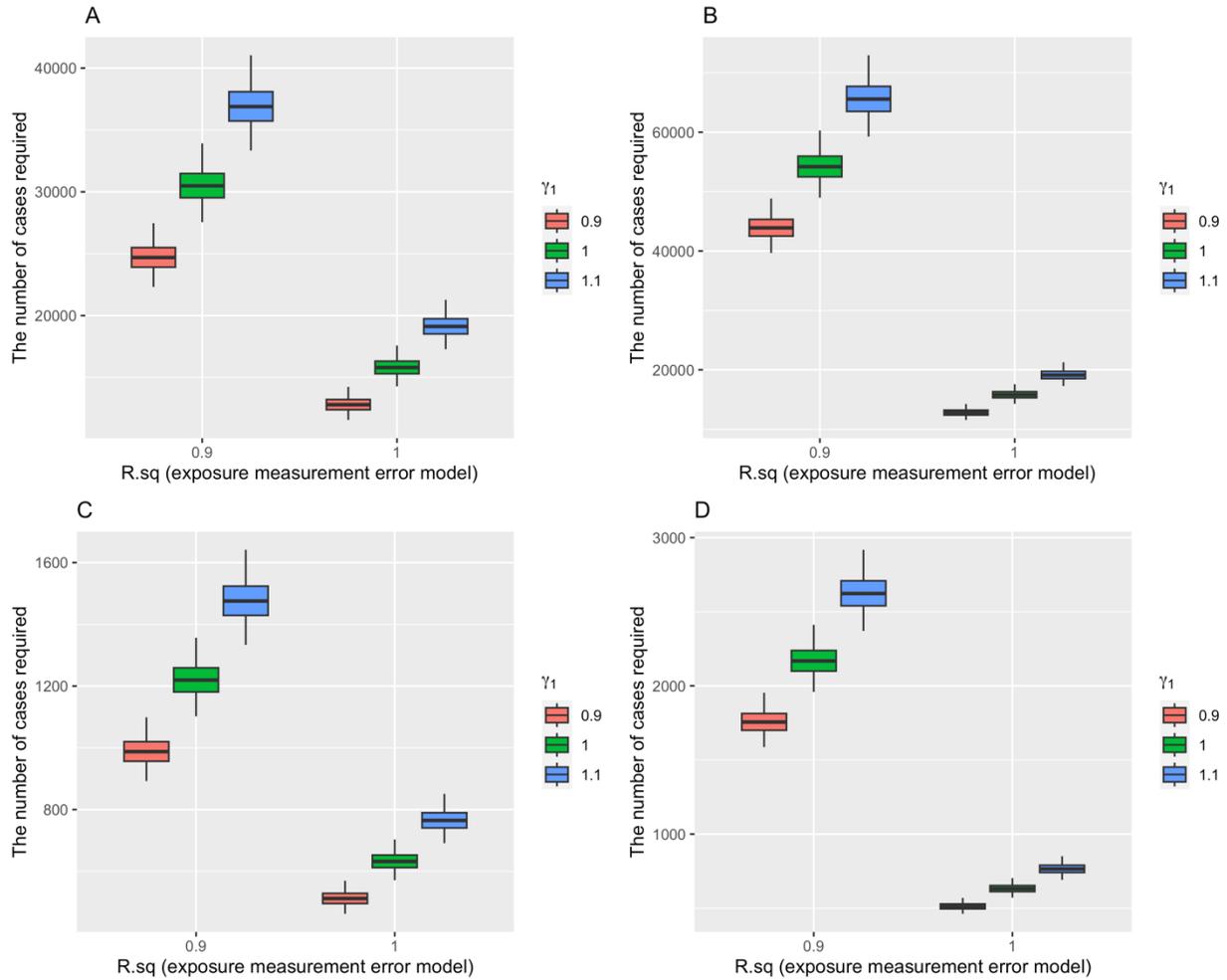

**Figure 3.** $n$ sensitiveness to $R^2_{x^{ep}_{g,t}, x_{i,t}}$ and $\gamma_1$ under the assumptions #3 (A and C) and #4 (B, D) and $\theta_l = 0.01$ (A and B) and $\theta_l = 0.05$ (C and D)

Note. When $R^2_{x^{ep}_{g,t}, x_{i,t}} = 1$, $\gamma_1 = 0.9$ would lead to overestimation where $g$ may or may not equal $i$. The results should be interpreted carefully.

**Web Appendix 1. Derivation of the standard error for the estimator in conditional logistic regression.**

**Web Appendix 2. Heteroscedastic variance calculations.**

**Web Appendix 3. Approximation for the estimators in the presence of additive linear (-like) error.**

**Web Appendix 4. Residual confounding in distributed lag models when the error is differential.**

**Web Appendix 5. Adjustment for differential errors.**

**Web Appendix 6. Standard error decomposition in the presence of multiplicative error.**

**Web Appendix 7. Heteroscedastic exposure variance approximation for multiplicative error.**

**Web Appendix 8. Approximation for the estimators without validation data in hand.**

**Web Appendix 9. Approximation for the estimator of a second-order polynomial.**

**Web Appendix 10. Simulation experiments.**

**Web Table 1. Linear regressions of measured $O_3$/$PM_{2.5}$ at EPA monitors against distributed temporal lags of estimated $O_3$/$PM_{2.5}$ (Berkson-like error perspective) by U.S. EPA's FAQSD models (Jun–Aug. 2020) and of estimated $O_3$/$PM_{2.5}$ by U.S. EPA's FAQSD models (Jun–Aug. 2020) against distributed temporal lags of measured $O_3$/$PM_{2.5}$ at EPA monitors (Linear-like error perspective)**

**Web Table 2. Empirical estimates of $\theta_l$ and $\bar{\theta}$ and various approximations for $\theta_l$ and $\bar{\theta}$ in simulation experiments**

**Web Table 3. Empirical standard error estimates and approximations in simulation experiments**

**Web Figure 1.** Relationship between daily estimated $O_3$/$PM_{2.5}$ and daily measured $O_3$/$PM_{2.5}$ at monitors in Rio de Janeiro, Brazil for the years 2012–2017 (A, B, E, and F) and in the contiguous United States (C, D, G, and H) for Jun.–Aug. 2020

**Web Figure 2.** Residual plots including partial autocorrelation plots for linear regressions in Table 1 of the main manuscript

**Web Figure 3.** Scatter plots for several sets of gold standard variables and variables measured with error created in simulation experiments

**Web Figure 4.** Power curves obtained from sample size calculation equations presented in the main text (dashed curves) and empirical statistical powers (points) from simulation experiments

**Web Appendix 1. Derivation of the standard error for the estimator in conditional logistic regression.**

Under the alternative hypothesis, $H_1$, the score equation, $U(\theta_l)$ for the coefficient for $x_{t-l}$ in conditional logistic regression, $\theta_l$, is

$$U(\theta_l) = \sum_{h=1}^{n} x_{h,1,l} - \frac{\sum_{j=1}^{m_h} x_{h,j,l} \exp(\theta_l x_{h,j,l})}{\sum_{j=1}^{m_h} \exp(\theta_l x_{h,j,l})}$$

The Fisher information is

$$I(\theta_l) = \sum_{h=1}^{n} \left[ \frac{\sum_{j=1}^{m_h} x_{h,j,l}^2 \exp(\theta_l x_{h,j,l})}{\sum_{j=1}^{m_h} \exp(\theta_l x_{h,j,l})} - \left( \frac{\sum_{j=1}^{m_h} x_{h,j,l} \exp(\theta_l x_{h,j,l})}{\sum_{j=1}^{m_h} \exp(\theta_l x_{h,j,l})} \right)^2 \right] = \sum_{h=1}^{n} V_{1h}(x_{t-l})$$

Under the null hypothesis, $H_0$,

$$U(\theta_l = 0) = \sum_{h=1}^{n} \left( x_{h,1,l} - \frac{\sum_{j=1}^{m_h} x_{h,j,l}}{m_h} \right) = \sum_{h=1}^{n} (x_{h,1,l} - \bar{x}_{h,l})$$

$$I(\theta_l = 0) = \sum_{h=1}^{n} \left[ \frac{\sum_{j=1}^{m_h} x_{h,j,l}^2}{m_h} - \left( \frac{\sum_{j=1}^{m_h} x_{h,j,l}}{m_h} \right)^2 \right]$$

$$= \sum_{h=1}^{n} \left[ \frac{\sum_{j=1}^{m_h} (x_{h,j,l} - \bar{x}_{h,l})^2}{m_h} \right] = \sum_{h=1}^{n} \sigma_h^2(x_{t-l})$$

where $\bar{x}_{h,l} = \frac{\sum_{j=1}^{m_h} x_{h,j,l}}{m_h}$. Using a Taylor's expansion, the equation for power analysis for a value of $\theta_l$ under $H_1$ becomes

$$|\theta_l| \sqrt{\sum_{h=1}^{n} \sigma_h^2(x_{t-l})} = Z_{1-\alpha} + Z_{1-\beta} \sqrt{\frac{\sum_{h=1}^{n} V_{1h}(x_{t-l})}{\sum_{h=1}^{n} \sigma_h^2(x_{t-l})}}$$

$$Z_{1-\beta} = \frac{|\theta_l| \sqrt{\sum_{h=1}^{n} \sigma_h^2(x_{t-l})} - Z_{1-\alpha}}{\sqrt{\frac{\sum_{h=1}^{n} V_{1h}(x_{t-l})}{\sum_{h=1}^{n} \sigma_h^2(x_{t-l})}}} \approx |\theta_l| \sqrt{\sum_{h=1}^{n} \sigma_h^2(x_{t-l})} - Z_{1-\alpha}$$

where $\sum_{h=1}^{n} V_{1h}(x_{t-l}) \approx \sum_{h=1}^{n} \sigma_h^2(x_{t-l})$ under a local alternative, Finally,

$$SE(\hat{\theta}_l) = \sqrt{I(\theta_l)^{-1}} = \sqrt{\left[\sum_{h=1}^{n} V_{1h}(x_{t-l})\right]^{-1}} \approx \sqrt{\left[\sum_{h=1}^{n} \sigma_h^2(x_{t-l})\right]^{-1}}$$

**Web Appendix 2. Heteroscedastic variance calculations.**

If health data are not available before the initiation of a study (e.g., investigators need to recruit patients, but their residential history is unknown before recruitment), investigators would not be able to exactly calculate $\sigma_h^2$. Instead, they may be able to have external information about a source population. They could approximately calculate $\sigma_h^2$ with the information about *k*-th sub-source populations.

Suppose that investigators have exposure data or information about the distribution of exposure in a source population, $\bar{\sigma}^2$ can be approximated to the expected case number-weighted $\sigma^2$ as

$$\bar{\sigma}^2 \approx \frac{\sum_{ss=1}^{SS} E[Cases_{ss}]\sigma_{ss}^2}{\sum_{ss=1}^{SS} E[Cases_{ss}]}$$

where $E[Cases_{ss}]$ is the expected number of cases in *ss*-th sub-source population, *SS* is the total number of sub-source populations, and $\sigma_{ss}^2$ is the variance of exposure in *ss*-th sub-source population. Note that controls must come from the same (sub-) source population to avoid selection bias. $\frac{\sum_{ss=1}^{SS} E[Cases_{ss}]\sigma_{ss}^2}{\sum_{ss=1}^{SS} E[Cases_{ss}]} \to \bar{\sigma}^2$ with the higher number of (self-)controls in matching.

Investigators may use external data such as population statistics or summary of electronic health records to identify the expected number of cases. One way to estimate $\bar{\sigma}^2$ is

$$\bar{\sigma}^2 \approx \frac{\sum_{fl=1}^{N_{fl}} Pop_{fl} \sum_{s=1}^{S} R_{fl,s}\sigma_{fl,s}^2}{\sum_{fl=1}^{N_{fl}} Pop_{fl} \sum_{s=1}^{S} R_{fl,s}}$$

where $Pop_{fl}$ denotes a sub-source population count for a fixed location $fl$. (e.g., the number of residents in a census tract $fl$). $N_{fl}$ is the total number of the sub-source populations for a study (e.g., the number of census tracts in a city). $\sigma^2_{fl,s}$ is the variance for $fl$ and matched stratum $s$ (i.e., matching each case to their self-control(s)). $S$ is the total number of the strata.

**Web Appendix 3. Approximation for the estimators in the presence of additive linear (-like) error.**

Suppose the correct linear outcome model, $E[Y_{i,t}] = \vartheta + \sum_{l=0}^{L} \theta_l x_{i,t-l} + \boldsymbol{\vartheta}\mathbf{z}$ and $g = i$. For small effect size or exposure measurement error, the consequence from linear regression may be applicable.

In practice, investigators may use $x_{g,t}^{ep}$ in place of $x_{i,t}$. Suppose a linear (-like) (L(L)) error model, $x_{g,t}^{ep} = \gamma_0 + \eta_0 x_{g,t} + u_{g,t}^{L(L)}$. (See Section 3.3). To consider the impact of this error model on effect estimators, it is convenient to consider following re-written error models based on regression calibration,

$$x_{g,t} = \gamma_{0,0}^* + \sum_{l=0}^{L} \eta_{0,l}^* x_{g,t-l}^{ep} + \boldsymbol{\eta}_{0,z}\mathbf{z} + u_{g,t}^*$$

$$x_{g,t-1} = \gamma_{1,0}^* + \sum_{l=0}^{L} \eta_{1,l}^* x_{g,t-l}^{ep} + \boldsymbol{\eta}_{1,z}\mathbf{z} + u_{g,t-1}^*$$

$$\ldots$$

$$x_{g,t-L} = \gamma_{L,0}^* + \sum_{l=0}^{L} \eta_{L,l}^* x_{g,t-l}^{ep} + \boldsymbol{\eta}_{L,z}\mathbf{z} + u_{g,t-L}^*$$

For $x_{g,t-l}^{ep}$, the following parameters are important:

$$\eta_{0,l}^* = \frac{Cov\left(x_{g,t}, x_{g,t-l}^{ep} \middle| \mathbf{x}_g^{ep-l}, \mathbf{z}\right)}{V\left(x_{g,t-l}^{ep} \middle| \mathbf{x}_g^{ep-l}, \mathbf{z}\right)} \cong \frac{\eta_0 Cov\left(x_{g,t}, x_{g,t-l} \middle| \mathbf{x}_g^{ep-l}, \mathbf{z}\right) + Cov\left(x_{g,t}, u_{g,t-l}^{L(L)} \middle| \mathbf{x}_g^{ep-l}, \mathbf{z}\right)}{\eta_0^2 V\left(x_{g,t-l} \middle| \mathbf{x}_g^{ep-l}, \mathbf{z}\right) + \eta_0 Cov\left(x_{g,t-l}, u_{g,t-l}^{L(L)} \middle| \mathbf{x}_g^{ep-l}, \mathbf{z}\right) + V\left(u_{g,t-l}^{L(L)} \middle| \mathbf{x}_g^{ep-l}, \mathbf{z}\right)}.$$

$$\eta_{1,l}^* = \frac{Cov\left(x_{g,t-1}, x_{g,t-l}^{ep} \middle| \mathbf{x}_g^{ep-l}, \mathbf{z}\right)}{V\left(x_{g,t-l}^{ep} \middle| \mathbf{x}_g^{ep-l}, \mathbf{z}\right)} \cong \frac{\eta_0 Cov\left(x_{g,t-1}, x_{g,t-l} \middle| \mathbf{x}_g^{ep-l}, \mathbf{z}\right) + Cov\left(x_{g,t-1}, u_{g,t-l}^{L(L)} \middle| \mathbf{x}_g^{ep-l}, \mathbf{z}\right)}{\eta_0^2 V\left(x_{g,t-l} \middle| \mathbf{x}_g^{ep-l}, \mathbf{z}\right) + \eta_0 Cov\left(x_{g,t-l}, u_{g,t-l}^{L(L)} \middle| \mathbf{x}_g^{ep-l}, \mathbf{z}\right) + V\left(u_{g,t-l}^{L(L)} \middle| \mathbf{x}_g^{ep-l}, \mathbf{z}\right)}.$$

…

$$\eta_{L,l}^* = \frac{Cov\left(x_{g,t-L}, x_{g,t-l}^{ep} | x_g^{ep-l}, z\right)}{V\left(x_{g,t-l}^{ep} | x_g^{ep-l}, z\right)} \cong \frac{\eta_0 Cov\left(x_{g,t-L}, x_{g,t-l} | x_g^{ep-l}, z\right) + Cov\left(x_{g,t-L}, u_{g,t-l}^{L(L)} | x_g^{ep-l}, z\right)}{\eta_0^2 V\left(x_{g,t-l} | x_g^{ep-l}, z\right) + \eta_0 Cov\left(x_{g,t-l}, u_{g,t-l}^{L(L)} | x_g^{ep-l}, z\right) + V\left(u_{g,t-l}^{L(L)} | x_g^{ep-l}, z\right)}.$$

Thus, generally,

$$\eta_{j,l}^* = \frac{\eta_0 Cov\left(x_{g,t-j}, x_{g,t-l} | x_g^{ep-l}, z\right) + Cov\left(x_{g,t-j}, u_{g,t-l}^{L(L)} | x_g^{ep-l}, z\right)}{\eta_0^2 V\left(x_{g,t-l} | x_g^{ep-l}, z\right) + \eta_0 Cov\left(x_{g,t-l}, u_{g,t-l}^{L(L)} | x_g^{ep-l}, z\right) + V\left(u_{g,t-l}^{L(L)} | x_g^{ep-l}, z\right)}$$

If the error is independent and identically distributed (independent of $x_{g,t}$) (i.e., non-differntial error), then $Cov\left(x_{g,t-j}, u_{g,t-l}^{L(L)} | x_g^{ep-l}, z\right) = 0$ so that

$$\eta_{j,l}^* = \frac{\eta_0 Cov\left(x_{g,t-j}, x_{g,t-l} | x_g^{ep-l}, z\right)}{\eta_0^2 V\left(x_{g,t-l} | x_g^{ep-l}, z\right) + V\left(u_{g,t-l}^* | x_g^{ep-l}, z\right)}$$

and

$$\eta_{l,l}^* = \frac{\eta_0 V\left(x_{g,t-l} | x_g^{ep-l}, z\right)}{\eta_0^2 V\left(x_{g,t-l} | x_g^{ep-l}, z\right) + V\left(u_{g,t-l}^* | x_g^{ep-l}, z\right)} \equiv \lambda_l$$

where $\lambda_l$ is a bias factor.

The correct linear outcome model can be re-expressed as

$$E[Y_{i,t}] = \vartheta + \sum_{l=0}^{L} \theta_l (\gamma_{l,0}^* + \sum_{j=0}^{L} \eta_{l,j}^* x_{g,t-j}^{ep} + \boldsymbol{\eta}_{l,z} \boldsymbol{z} + u_{g,t-l}^*) + \boldsymbol{\vartheta z}$$

But, investigators use

$$E[Y_{i,t-l}] = \vartheta^* + \sum_{l=0}^{L} \theta_l^{L(L)} x_{g,t-l}^{ep} + \boldsymbol{\vartheta^* z}$$

If $\eta_{l,j}^* = 0$ for $j \neq l$, which occurs when $Cov\left(x_{g,t-j}, x_{g,t-l} | x_g^{ep-l}, z\right) = 0$ (i.e., no autocorrelation conditional on $x_g^{ep-l}, z$)

$$\theta_l^{L(L)} \cong \theta_l \eta_{l,l}^* + \sum_{k=0}^{L} \theta_k \varrho_{u_{g,t-k}^*, x_{g,t-l}^{ep} | x_{g,t}^{ep-l}, z}$$

If $u_{g-k}^*$ is independent and identically distributed (and independent of $x_{g,t-l}^{ep}$) given $x_{g,t}^{ep-l}$ and $z$, which may occur when the error is non-differential, $\varrho_{u_{g,t-k}^*, x_{g,t-l}^{ep} | x_{g,t}^{ep-l}, z} = 0$. Then,

$$\theta_l^L \cong \theta_l \eta_{l,l}^* = \theta_l \lambda_l$$

If $\eta_{l,j}^* \neq 0$ (e.g., non-zero correlations between lagged variables of $x_{g,t}$) and/or $u_{g-k}^*$ is not independent of $x_{g,t-l}^{ep}$ even conditional on $x_{g,t}^{ep-l}$ and $z$ ($\varrho_{u_{g,t-k}^*, x_{g,t-l}^{ep} | x_{g,t}^{ep-l}, z} \neq 0$), then the estimator would vary. Recall the following components in the re-written correct outcome model

$$\sum_{j=0}^{L} \theta_0 \eta_{0,j}^* x_{g,t-j}^{ep} = \theta_0 \eta_{0,0}^* x_{g,t}^{ep} + \theta_0 \eta_{0,1}^* x_{g,t-1}^{ep} + \cdots + \theta_0 \eta_{0,L}^* x_{g,t-L}^{ep} \text{ (for } l=0\text{)}$$

$$\sum_{j=0}^{L} \theta_1 \eta_{1,0,j}^* x_{g,t-j}^{ep} = \theta_1 \eta_{1,0}^* x_{g,t}^{ep} + \theta_1 \eta_{1,1}^* x_{g,t-1}^{ep} + \cdots + \theta_1 \eta_{1,L}^* x_{g,t-L}^{ep} \text{ (for } l=1\text{)}$$

…

$$\sum_{j=0}^{L} \theta_L \eta_{L,0}^* x_{g,t-j}^{ep} = \theta_L \eta_{L,0}^* x_{g,t}^{ep} + \theta_L \eta_{L,1}^* x_{g,t-1}^{ep} + \cdots + \theta_L \eta_{L,L}^* x_{g,t-L}^{ep} \text{ (for } l=L\text{)}$$

so that $\theta_l^{L(L)}$ can be approximated as

$$\theta_l^{L(L)} \cong \sum_{j=0}^{L} \theta_j \eta_{j,l}^* + \sum_{k=0}^{L} \theta_k \varrho_{u_{g,t-k}^*, x_{g,t-l}^{ep} | x_{g,t}^{ep-l}, z}$$

This completes the development of an approximation equation for $\theta_l^{L(L)}$. Below is an investigation of the behavior of $\theta_l^{L(L)}$.

The approximation equation above shows that the following two merit investigations:

1) How $\theta_l^{L(L)}$ for $j \neq l$ depends on $\eta_{j,l}^*$ and $\varrho_{u_{g,t-k}^*, x_{g,t-l}^{ep} | x_{g,t}^{ep-l}, \mathbf{z}}$

2) The covariance component in the numerator of $\eta_{j,l}^*$, which is

$$Cov\left(x_{g,t-j}, u_{g,t-l}^{L(L)} \middle| \mathbf{x}_g^{ep-l}, \mathbf{z}\right)$$

For 1), when $x_{g,t}$ is autocorrelated (e.g., $x_{g,t} = \zeta_1 x_{g,t-1} + \zeta_2 x_{g,t-2} + \cdots$) and $\mathbf{x}_g^{ep-l}, \mathbf{z}$ are not sufficient to explain this autocorrelation, then, for some $j$ ($j \neq l$), $Cov\left(x_{g,t-j}, x_{g,t-l} \middle| \mathbf{x}_g^{ep-l}, \mathbf{z}\right) \neq 0$ so that $\eta_{l,j}^* \neq 0$. Note that by the definition, if $x_{g,t-l}$ has an effect on $Y$ and is correlated with $x_{g,t-j}$, this would be seen as a confounder. Residual confounding would likely occur if the error in $x_{g,t}^{ep}$ is not small because if this is not small, $\mathbf{x}_g^{ep-l}, \mathbf{z}$ would be unlikely to be sufficient to make $Cov\left(x_{g,t-j}, x_{g,t-l} \middle| \mathbf{x}_g^{ep-l}, \mathbf{z}\right) = 0$ so that $\eta_{l,j}^* \neq 0$. $\varrho_{u_{g,t-k}^*, x_{g,t-l}^{ep} | x_{g,t}^{ep-l}, \mathbf{z}}$ may be negligible, depending on the structure of $x_{g,t}, x_{g,t}^{ep}, \mathbf{z}$.

For 2), $Cov\left(x_{g,t-j}, u_{g,t-l}^{L(L)} \middle| \mathbf{x}_g^{ep-l}, \mathbf{z}\right)$ may be zero if $u_{g,t}^L$ is completely random. Otherwise, this may not be zero.

**Web Appendix 4. Residual confounding in distributed lag models when the error is differential.**

Recall

$$\theta_l^{L(L)} \cong \sum_{j=0}^{L} \theta_j \eta_{j,l}^* + \sum_{k=0}^{L} \theta_k \varrho_{u_{g,t-k}^*, x_{g,t-l}^{ep} | x_{g,t}^{ep-l}, z}$$

and

$$Cov\left(x_{g,t-j}, u_{g,t-l}^{L(L)} \middle| x_g^{ep-l}, z\right) \text{ in } \eta_{j,l}^*$$

in Web Appendix 3. These may be applicable for non-linear models such as logistic regression when the effect size is small, or the measurement error is small.

If the error component is $u_{g,t-l}^{LL}$, it is convenient to distinguish between

$Cov\left(x_{g,t-l}, u_{g,t-l}^{LL} \middle| x_g^{ep-l}, z\right)$ (i.e., $j = l$) and $Cov\left(x_{g,t-j}, u_{g,t-l}^{LL} \middle| x_g^{ep-l}, z\right)$ for $j \neq l$.

Let me re-investigate 2) in Web Appendix 3 to examine the behavior of $\theta_l^{L(L)}$ in more detail.

2-1) If $Cov\left(x_{g,t-l}, u_{g,t-l}^{LL} \middle| x_g^{ep-l}, z\right) \neq 0$, then $\eta_{l,l}^* \neq \lambda$.

This is a consequence when adjustment for $x_g^{ep-l}, z$ is not adequate to eliminate the correlation between $x_{g,t-l}$ and $u_{g,t-l}^{LL}$. This means that the exposure measurement error would be still differential even after conditioning on $x_g^{ep-l}, z$ because the correlation between $u_{g,t-l}^{LL}$ and $Y$ would be non-zero through a non-zero correlation between $x_{g,t-l}$ and $u_{g,t-l}^{LL}$ given $x_g^{ep-l}, z$ if $X$ has an effect on $Y$. Recall that the definition of differential exposure measurement error is that the error is not independent of $Y$. To understand the behavior of $\theta_l^{L(L)}$, suppose $\eta_0 > 0$, which would be realistic in epidemiological investigations because we would otherwise not use $x^{ep}$ at all. Then,

$$\eta^{*}_{l,l} = \frac{\eta_0 V\left(x_{g,t-l}|x_g^{ep-l}, z\right) + Cov\left(x_{g,t-l}, u_{g,t-l}^{LL}\Big|x_g^{ep-l}, z\right)}{V\left(x_{g,t-l}^{ep}|x_g^{ep-l}, z\right)}$$

$$= \frac{\eta_0 V\left(x_{g,t-l}|x_g^{ep-l}, z\right) + \dfrac{Cov\left(x_{g,t-l}, u_{g,t-l}^{LL}\Big|x_g^{ep-l}, z\right) V\left(x_{g,t-l}|x_g^{ep-l}, z\right)}{V\left(x_{g,t-l}|x_g^{ep-l}, z\right)}}{V\left(x_{g,t-l}^{ep}|x_g^{ep-l}, z\right)}$$

$$= \frac{\eta_0 V\left(x_{g,t-l}|x_g^{ep-l}, z\right) + \varrho_{u_{g,t-l}^{LL}, x_{g,t-l}|x_g^{ep-l}, z} V\left(x_{g,t-l}|x_g^{ep-l}, z\right)}{V\left(x_{g,t-l}^{ep}|x_g^{ep-l}, z\right)}$$

$$= \frac{(\eta_0 + \varrho_{u_{g,t-l}^{LL}, x_{g,t-l}|x_g^{ep-l}, z}) V\left(x_{g,t-l}|x_g^{ep-l}, z\right)}{V\left(x_{g,t-l}^{ep}|x_g^{ep-l}, z\right)}$$

$$= \frac{(\eta_0 + \varrho_{u_{g,t-l}^{LL}, x_{g,t-l}|x_g^{ep-l}, z}) V\left(x_{g,t-l}|x_g^{ep-l}, z\right)}{\eta_0^2 V\left(x_{g,t-l}|x_g^{ep-l}, z\right) + \eta_0 Cov\left(x_{g,t-l}, u_{g,t-l}^{LL}|x_g^{ep-l}, z\right) + V\left(u_{g,t-l}^{LL}\Big|x_g^{ep-l}, z\right)}$$

$$= \frac{(\eta_0 + \varrho_{u_{g,t-l}^{LL}, x_{g,t-l}|x_g^{ep-l}, z})}{\eta_0\left(\eta_0 + \varrho_{u_{g,t-l}^{LL}, x_{g,t-l}|x_g^{ep-l}, z}\right) + \dfrac{V\left(u_{g,t-l}^{LL}\Big|x_g^{ep-l}, z\right)}{V\left(x_{g,t-l}|x_g^{ep-l}, z\right)}}$$

$$= \frac{1}{\eta_0 + \dfrac{V\left(u_{g,t-l}^{LL}\Big|x_g^{ep-l}, z\right)}{V\left(x_{g,t-l}|x_g^{ep-l}, z\right)\left(\eta_0 + \varrho_{u_{g,t-l}^{LL}, x_{g,t-l}|x_g^{ep-l}, z}\right)}}$$

If the error is small so that $\frac{V\left(u_{g,t-l}^{LL}|x_g^{ep-l},z\right)}{V\left(x_{g,t-l}|x_g^{ep-l},z\right)} \approx 0$, then $\frac{V\left(u_{g,t-l}^{LL}|x_g^{ep-l},z\right)}{V\left(x_{g,t-l}|x_g^{ep-l},z\right)\left(\eta_0+\varrho_{u_{g,t-l}^{LL},x_{g,t-l}|x_g^{ep-l},z}\right)} \approx 0$,

then finally,

$$\eta_{l,l}^* \approx \frac{1}{\eta_0}$$

Or, if the error is not small, let $a = \frac{V\left(u_{g,t-l}^{LL}|x_g^{ep-l},z\right)}{V\left(x_{g,t-l}|x_g^{ep-l},z\right)\left(\eta_0+\varrho_{u_{g,t-l}^{LL},x_{g,t-l}|x_g^{ep-l},z}\right)}$, then

$$\eta_{l,l}^* \approx \frac{1}{\eta_0 + a}$$

If the error is large so that $a \geq 1$, then $0 < \eta_{l,l}^* < 1$.

If $\eta_0 \geq 1$, then $0 < \eta_{l,l}^* \leq 1$, which is similar to the consequence by $\lambda_l$ for non-differential linear error. To compare this with $\lambda_l$,

$$\lambda_l = \frac{\eta_0 V\left(x_{g,t-l}|x_g^{ep-l},z\right)}{V\left(x_{g,t-l}^{ep}|x_g^{ep-l},z\right)} = \frac{1}{\eta_0 + \frac{V\left(u_{g,t-l}|x_g^{ep-l},z\right)}{V\left(x_{g,t-l}|x_g^{ep-l},z\right)(\eta_0)}}$$

If $x_{g,t-l}$ is autocorrelated, then $x_{g,t-l}$ would also be correlated with $x_g^{ep-l}$ unless the error is not very large. This implies that $\varrho_{u_{g,t-l}^{LL},x_{g,t-l}|x_g^{ep-l},z}$ may not be high because $x_g^{ep-l}$ is also correlated with $u_{t-l}$ so that the correlation between $u_{g,t-l}^{LL}$ and $x_{g,t-l}$ may be adjusted for to some degree by conditioning on $x_g^{ep-l}, z$. Thus, it is reasonable to conclude that $\eta_{l,l}^*$ may act like $\lambda_l$ of non-differential linear error in distributed lag models if the error is not large.

2-2) Regarding $Cov\left(x_{g,t-j}, u_{g,t-l}^{LL} \mid x_g^{ep-l}, z\right) \neq 0$ for $j \neq l$.

This term subsumes the situation that the exposure measurement error is correlated with $x_{g,t-j}$ after adjustment for $x_g^{ep-l}, z$ was performed. There are two things to consider:

Q1. Can $x_{g,t-j}$ is correlated with $u_{g,t-l}^{LL}$?

A1. If $x_{g,t}$ is autocorrelated and $u_{g,t-l}^{LL}$ is autocorrelated (e.g., because both have patterns over space, time, or both). Recall that $u_{g,t-l}^{LL}$ may include $\sum_{j=1}^{J} \eta_j x_{g,t-j}$ (See the main text) where $\eta_j$ is non-zero. So, it would be difficult to say that these two are not correlated.

Q2. If $x_{g,t-j}$ is correlated with $u_{g,t-l}^{LL}$, can this correlation be adjusted for using $x_{g,t-j}^{ep}$?

Adjusting for $x_{g,t-j}^{ep}$ implies that adjusting for $x_{g,t-j}$ and $u_{g,t-j}^{LL}$ simultaneously using one variable, but not separately (Recall a linear-like error model: $x_{g,t-j}^{ep} = \gamma_0 + \gamma_1 x_{g,t-j} + u_{g,t-j}^{LL}$). If the spatiotemporal pattern of $x_{g,t-j}$ and $u_{g,t-j}^{LL}$ would be similar, they would be highly correlated, then the performance of adjustment for $x_{g,t-j}^{ep}$ in terms of adjustment for $x_{g,t-j}$ would be substantial. Furthermore, the performance of adjustment for $x_{g,t-j}^{ep}$ in terms of adjustment for $u_{g,t-l}^{LL}$ would also be substantial. If the pattern of $x_{g,t-j}$ and $u_{g,t-j}^{LL}$ would not be similar, then these two would not be highly correlated, then the performance of adjustment for $x_{g,t-j}^{ep}$ would not be substantial. The extreme case is a non-differential error model (e.g., $x_{g,t-j}^{ep} = \gamma_0 + \gamma_1 x_{g,t-j} + u_{g,t-j}^{L}$ where $u_{g,t-j}^{L}$ is completely random).

These suggest that if the exposure measurement error is differential in a way that $u_{g,t-j}^{LL}$ is highly correlated with $x_{g,t-j}$, it would not be a stretch to say that $Cov\left(x_{g,t-j}, u_{g,t-l}^{LL} \middle| x_g^{ep-l}, z\right)$ may be negligible due to conditioning on $x_g^{ep-l}, z$.

If $Cov\left(x_{g,t-j}, u_{g,t-l}^{LL} \middle| x_g^{ep-l}, z\right)$ for $j \neq l$ is negligible (i.e., $\approx 0$) due to the high correlation between $x_{g,t-j}$ and $u_{g,t-l}^{LL}$, the term, $Cov\left(x_{g,t-j}, x_{g,t-l} \middle| x_g^{ep-l}, z\right)$ in $\eta_{j,l}^*$ may also be negligible due to conditioning on $x_g^{ep-l}, z$.

Recall

$$\eta_{j,l}^* = \frac{\eta_0 Cov\left(x_{g,t-j}, x_{g,t-l} \middle| x_g^{ep-l}, z\right) + Cov\left(x_{g,t-j}, u_{g,t-l}^{L(L)} \middle| x_g^{ep-l}, z\right)}{\eta_0^2 V\left(x_{g,t-l} \middle| x_g^{ep-l}, z\right) + \eta_0 Cov\left(x_{g,t-l}, u_{g,t-l}^{L(L)} \middle| x_g^{ep-l}, z\right) + V\left(u_{g,t-l}^{L(L)} \middle| x_g^{ep-l}, z\right)}$$

in Web Appendix 3. If $Cov\left(x_{g,t-j}, x_{g,t-l} \middle| x_g^{ep-l}, z\right) \approx 0$ and $Cov\left(x_{g,t-j}, u_{g,t-l}^{LL} \middle| x_g^{ep-l}, z\right) \approx 0$ for $j \neq l$, then,

$$\eta_{j,l}^* \approx 0$$

Recall,

$$\theta_l^{L(L)} \cong \sum_{j=0}^{L} \theta_j \eta_{j,l}^* + \sum_{k=0}^{L} \theta_k \varrho_{u_{g,t-k}^*, x_{g,t-l}^{ep} \middle| x_{g,t}^{ep-l}, z}$$

and $\eta_{j,l}^* \approx 0$ for $j \neq l$ implies

$$\theta_l^{L(L)} \approx \theta_l \eta_{l,l}^* + \sum_{k=0}^{L} \theta_k \varrho_{u_{g,t-k}^*, x_{g,t-l}^{ep} \middle| x_{g,t}^{ep-l}, z}$$

and $\varrho_{u_{g,t-k}^*, x_{g,t-l}^{ep} \middle| x_{g,t}^{ep-l}, z}$ may also become negligible if $x_{g,t}^{ep-l}$ is correlated with $x_{g,t-l}^{ep}$, then,

$$\theta_l^{L(L)} \approx \theta_l \eta_{l,l}^*$$

In 2-1), it was revealed that $\eta_{l,l}^*$ may act similar to $\lambda_l$.

In conclusion, if the additive error is autocorrelated such as additive LL error, the risk of the residual confounding due to the use of $x_g^{ep-l}$ may be smaller than what the risk would have been if the error was non-differential additive linear error.

**Web Appendix 5. Adjustment for differential errors.**

Before discussing adjustment for differential errors, recall the approximation proposed in Web Appendix 3,

$$\theta_l^{L(L)} \cong \sum_{j=0}^{L} \theta_j \eta_{j,l}^* + \sum_{k=0}^{L} \theta_k \varrho_{u_{g,t-k}^*, x_{g,t-l}^{ep} | x_{g,t}^{ep^{-l}}, z}$$

this may be applicable for non-linear models such as logistic regression when the effect size is small, or the measurement error is small. This approximation may be applicable for many error models in the form of $x_{g,t}^{ep} = g(x_{g,t}, v, u^{(A/M)L(L)})$ because this approximation does not rely on error models, but on re-written error models using regression calibration (Web Appendix 3), where $u^{L(L)}$ is non-differential error term, (including the autoregressive error model and multiplicative errors in the main manuscript) and $g$ is a transformation function, $v$ is an additional component describing the exposure measurement error (e.g., $x_{g,t-j}$ or $ð(Q^e)$). Simulation analyses in Web Appendix S10 demonstrate that this approximation equation performs well.

Regarding adjustment for differential errors, $v$ could be subsumed into $z$. As shown above and discussed in Web Appendix 3 and S4, the magnitude of the components in the approximation varies by $x_{g,t}^{ep^{-l}}, z$. Thus, if $z$ includes some variables that are related to the error, the contribution of the error to $\eta_{j,l}^*$ and $\varrho_{u_{g,t-k}^*, x_{g,t-l}^{ep} | x_{g,t}^{ep^{-l}}, z}$ may decrease. If all components that make error differential are controlled for, the remaining conditional error would become non-differential, which is referred to as conditional non-differential error.

**Web Appendix 6. Standard error decomposition in the presence of multiplicative error.**

Consider the variance decomposition,

$$SE\left(\hat{\theta}_l^{ML(L)}|\mathbf{z}, \mathbf{x}_g^{ep-l}\right)^2$$

$$= E\left[SE\left(\hat{\theta}_l^{ML(L)}|\exp(\gamma_{m0})\exp(u_{g,t}^{L(L)}), \mathbf{x}_g^{ep-l}, \mathbf{z}\right)^2\right]$$

$$+ V(E\left[\hat{\theta}_l^{ML(L)}|\exp(\gamma_{m0})\exp(u_{g,t}^{L(L)}), \mathbf{x}_g^{ep-l}, \mathbf{z}\right])$$

$$E\left[SE\left(\hat{\theta}_l^{ML(L)}|\exp(\gamma_{m0})\exp(u_{g,t}^{L(L)}), \mathbf{x}_g^{ep-l}, \mathbf{z}\right)^2\right] \cong \left(\sum_{h=1}^{n} V_{0h}((x_{g,t-l})^{\gamma_{m1}})\left(1 - R^2_{x_{g,t-l}^{ep}|\mathbf{x}_g^{ep-l},\mathbf{z}}\right)\right)^{-1}$$

$$V\left(E\left[\hat{\theta}_l^{ML(L)}|\exp(\gamma_{m0})\exp(u_{g,t}^{L(L)}), \mathbf{x}_g^{ep-l}, \mathbf{z}\right]\right)$$

$$\cong \theta_l^{ML(L)\,2}\left(\sum_{h=1}^{n} V_{0h}\left(\exp(\gamma_{m0})\exp(u_{g,t}^{L(L)})\right)\left(1 - R^2_{x_{g,t-l}^{ep}|\mathbf{x}_g^{ep-l},\mathbf{z}}\right)\right)^{-1}$$

and,

$$n\bar{\sigma}^2_{x_{g,t-l}^{ML(L)}} = \sum_{h=1}^{n} V_{0h}((x_{g,t-l})^{\gamma_{m1}})$$

$$n\bar{\sigma}^2_{\exp(\gamma_{m0})\exp(u_{i,t}^{L(L)})} = \sum_{h=1}^{n} V_{0h}\left(\exp(\gamma_{m0})\exp(u_{g,t}^{L(L)})\right)$$

**Web Appendix 7. Heteroscedastic exposure variance approximation for multiplicative error.**

Using the Delta method,

$$V(\gamma_{m1}\log(x_{g,t-l})) \approx \frac{\gamma_{m1}^2 V(x_{g,t-l})}{E[x_{g,t-l}]^2}$$

$$V((x_{g,t-l})^{\gamma_{m1}}) \approx \gamma_{m1}^2 E[x_{g,t-l}]^{2(\gamma_{m1}-1)} V(x_{g,t-l})$$

Recall $R^2_{\log(x^{ep}_{g,t-l}),\log(x_{g,t-l})} = \frac{V(\gamma_{m0}+\gamma_{m1}\log(x_{g,t-l}))}{V(\log(x^{ep}_{g,t-l}))} = \frac{V(\gamma_{m1}\log(x_{g,t-l}))}{V(\log(x^{ep}_{g,t-l}))}$. So,

$$V((x_{g,t-l})^{\gamma_{m1}}) \approx E[x_{g,t-l}]^{2\gamma_{m1}} V(\gamma_{m1}\log(x_{g,t-l}))$$

$$= E[x_{g,t-l}]^{2\gamma_{m1}} R^2_{\log(x^{ep}_{g,t-l}),\log(x_{g,t-l})} V(\log(x^{ep}_{g,t-l}))$$

Since $\bar{\sigma}^2_{x^{ML(L)}_{g,t-l}} = \frac{\sum_{h=1}^n \sigma_h^2(x^{ML(L)}_{g,t-l})}{n}$, $\bar{\sigma}^2_{\log(x^{ep}_{g,t-l})} = \frac{\sum_{h=1}^n \sigma_h^2(\log(x^{ep}_{g,t-l}))}{n}$,

$\sigma_h^2(x^{ML(L)}_{g,t-l}) \approx E[x_{h,g,l}]^{2\gamma_{m1}} R^2_{\log(x^{ep}_{g,t-l}),\log(x_{g,t-l})} \sigma_h^2\left(\log(x^{ep}_{g,t-l})\right)$, and

$\frac{\sum_{h=1}^n E[x_{h,g,l}]^{2\gamma_{m1}}}{n} \approx E[x_{g,t-l}]^{2\gamma_{m1}}$ for $\gamma_{m1} > 0$,

$$\bar{\sigma}^2_{x^{ML(L)}_{g,t-l}} \approx \bar{\sigma}^2_{\log(x^{ep}_{g,t-l})} E[x_{g,t-l}]^{2\gamma_{m1}} R^2_{\log(x^{ep}_{g,t-l}),\log(x_{g,t-l})}$$

# Web Appendix 8. Approximation for the estimators without validation data in hand.

Recall

$$\eta_{j,l}^* = \frac{Cov\left(x_{g,t-j}, x_{g,t-l}^{ep} \mid \boldsymbol{x}_g^{ep-l}, \boldsymbol{z}\right)}{V\left(X_{g,t-l}^{ep} \mid \boldsymbol{x}_g^{ep-l}, \boldsymbol{z}\right)}$$

$$= \frac{\eta_0 Cov\left(x_{g,t-j}, x_{g,t-l} \mid \boldsymbol{x}_g^{ep-l}, \boldsymbol{z}\right) + Cov\left(x_{g,t-j}, u_{g,t-l}^{L(L)} \mid \boldsymbol{x}_g^{ep-l}, \boldsymbol{z}\right)}{\eta_0^2 V\left(x_{g,t-l} \mid \boldsymbol{x}_g^{ep-l}, \boldsymbol{z}\right) + \eta_0 Cov\left(x_{g,t-l}, u_{g,t-l}^{L(L)} \mid \boldsymbol{x}_g^{ep-l}, \boldsymbol{z}\right) + V\left(u_{g,t-l}^{L(L)} \mid \boldsymbol{x}_g^{ep-l}, \boldsymbol{z}\right)}$$

in Appendices S2 and S3. This may be applicable for non-linear models such as logistic regression when the effect size is small, or the measurement error is small.

For $j \neq l$, without knowing $x_{g,t-j}$, we may use $x_{g,t-j}^{ep}$ to approximate this quantity. For illustration, consider $V(x_{g,t-j} \mid \boldsymbol{x}_g^{ep-l}, \boldsymbol{z})$. Using the notation in Web Appendix 3, recall

$$x_{g,t-j} = \gamma_{j,0}^* + \sum_{l=0}^{L} \eta_{j,l}^* x_{g,t-l}^{ep} + \boldsymbol{\eta}_{j,z} \boldsymbol{z} + u_{g,t-j}^*$$

Assume $V(x_{g,t-j} \mid \boldsymbol{x}_g^{ep-l}, \boldsymbol{z}) \approx V(x_{g,t-j} \mid \boldsymbol{x}_g^{ep}, \boldsymbol{z}) = V(u_{g,t-j}^*)$ so that

$Cov\left(x_{g,t-j}, x_{g,t-l} \mid \boldsymbol{x}_g^{ep-l}, \boldsymbol{z}\right) \approx V(u_{g,t-j}^*)$, which will be discussed below. Then,

$$\eta_{j,l}^* \approx \frac{\eta_0 V(u_{g,t-j}^*) + Cov\left(x_{g,t-j}, u_{g,t-l}^{L(L)} \mid \boldsymbol{x}_g^{ep-l}, \boldsymbol{z}\right)}{V\left(X_{g,t-l}^{ep} \mid \boldsymbol{x}_g^{ep-l}, \boldsymbol{z}\right)}$$

Note

$$R^2_{x_{g,t-l}^{ep}, x_{g,t-l} \mid \boldsymbol{x}_g^{ep-l}, \boldsymbol{z}} = \frac{V(\gamma_0 + \eta_0 x_{g,t-l} \mid \boldsymbol{x}_g^{ep-l}, \boldsymbol{z})}{V(x_{g,t-l}^{ep} \mid \boldsymbol{x}_g^{ep-l}, \boldsymbol{z})}$$

In the main manuscript, the notation $\gamma_1$ is used in place of $\eta_0$ while $\eta_0$ is used in Web Appendix 3 because this notation is used to highlight LL error models introduced in Section 3.3.

This can be further extended to

$$R^2_{x^{ep}_{g,t-l}, x_{g,t-l} | x^{ep-l}_g, z} \approx \frac{V\left(\gamma_0 + \eta_0 x_{g,t-l} \middle| x^{ep-l}_g, z\right)}{V\left(x^{ep}_{g,t-l} \middle| x^{ep-l}_g, z\right)} = \frac{\eta_0^2 V\left(x_{g,t-l} \middle| x^{ep-l}_g, z\right)}{V\left(x^{ep}_{g,t-l} \middle| x^{ep-l}_g, z\right)}$$

$$\approx \frac{\eta_0^2 \left(V(u^*_{g,t}) + V(\eta^*_{l,l} x^{ep}_{g,t-l} | x^{ep-l}_g, z)\right)}{V\left(x^{ep}_{g,t-l} \middle| x^{ep-l}_g, z\right)} = \frac{\eta_0^2 \left(V(u^*_{g,t}) + \eta^{*2}_{l,l} V\left(x^{ep}_{g,t-l} \middle| x^{ep-l}_g, z\right)\right)}{V\left(x^{ep}_{g,t-l} \middle| x^{ep-l}_g, z\right)}$$

$$= \frac{\eta_0^2 V(u^*_{g,t})}{V\left(x^{ep}_{g,t-l} \middle| x^{ep-l}_g, z\right)} + \eta_0^2 \eta^{*2}_{l,l}$$

Now compare

$$R^2_{x^{ep}_{g,t-l}, x_{g,t-l} | x^{ep-l}_g, z} \approx \frac{\eta_0^2 V(u^*_{g,t})}{V\left(x^{ep}_{g,t-l} \middle| x^{ep-l}_g, z\right)} + \eta_0^2 \eta^{*2}_{l,l}$$

$$\eta^*_{j,l} \approx \frac{\eta_0 V(u^*_{g,t-j}) + Cov\left(x_{g,t-j}, u^{L(L)}_{g,t-l} \middle| x^{ep-l}_g, z\right)}{V\left(X^{ep}_{g,t-l} \middle| x^{ep-l}_g, z\right)}$$

Appendices S2–S4 discusses when $Cov\left(x_{g,t-j}, u^{L(L)}_{g,t-l} \middle| x^{ep-l}_g, z\right) \approx 0$. So, it is possible

$$\eta^*_{j,l} \approx \frac{\eta_0 V(u^*_{g,t-j})}{V\left(X^{ep}_{g,t-l} \middle| x^{ep-l}_g, z\right)}$$

If $R^2_{x^{ep}_{g,t}, x_g} \approx R^2_{x^{ep}_{g,t-l}, x_{g,t-l} | x^{ep-l}_g, z}$, which is likely when $z$ is not a strong determinant of $x_g$,

$$R^2_{x^{ep}_{g,t}, x_g} \approx \eta_0 \eta^*_{j,l} + \eta_0^2 \eta^{*2}_{l,l}$$

For $\eta^*_{l,l}$, see Web Appendix 3 or recall 2-1 in Web Appendix 4. If $\eta^*_{l,l} \approx \lambda \approx \lambda_l$

where $\lambda = \dfrac{\eta_0 V\left(x_{g,t}|x_g^{ep-0},z\right)}{\eta_0^2 V\left(x_{g,t}|x_g^{ep-0},z\right)+V\left(u_{g,t}|x_g^{ep-0},z\right)}$,

$$R^2_{x_{g,t}^{ep},x_g} - \eta_0^2 \lambda^2 \approx \eta_0 \eta^*_{j,l}$$

and,

$$\eta_0 \lambda = \dfrac{V\left(\gamma_0 + \eta_0 x_{g,t}|x_g^{ep-0},z\right)}{\eta_0^2 V\left(x_{g,t}|x_g^{ep-0},z\right) + V\left(u_{g,t}|x_g^{ep-0},z\right)} = R^2_{x_{g,t}^{ep},x_{g,t}|x_g^{ep0},z}$$

Recall

$$\theta_l^{L(L)} \cong \sum_{j=0}^{L} \theta_j \eta^*_{j,l} + \sum_{k=0}^{L} \theta_k \varrho_{u^*_{g,t-k}, x^{ep}_{g,t-l}|x^{ep-l}_{g,t},z}$$

Assuming $\varrho_{u^*_{g,t-k}, x^{ep}_{g,t-l}|x^{ep-l}_{g,t},z} = 0$,

$$\theta_l^L \cong \sum_{j=0}^{L} \theta_j \eta^*_{j,l}$$

Finally,

$$\theta_l^{L(L)} \approx R^2_{x^{ep}_{g,t},x_g} \left( \dfrac{\theta_l + \left(1 - R^2_{x^{ep}_{g,t},x_g}\right)\sum_{j \neq l}^{L} \theta_j}{\eta_0} \right)$$

This approximation would perform reasonably if the assumption $Cov\left(x_{g,t-j}, x_{g,t-l}|x_g^{ep-l},z\right) \approx V(u^*_{g,t-j})$ is reasonable. Importantly, $V(u^*_{g,t-j})$ is only positive so that that approximation would

perform reasonably if there were no severe negative residual confounding. In practice, $Cov\left(x_{g,t-j}, x_{g,t-l} | x_g^{ep-l}, z\right)$ can be negative so that

$$\eta_{j,l}^* \approx \frac{\eta_0 Cov\left(x_{g,t-j}, x_{g,t-l} | x_g^{ep-l}, z\right)}{V\left(X_{g,t-l}^{ep} | x_g^{ep-l}, z\right)} < 0$$

which is not considered in the approximation.

To consider that, why and when could $Cov\left(x_{g,t-j}, x_{g,t-l} | x_g^{ep-l}, z\right)$ be negative so that $\eta_{j,l}^*$ be negative?

To answer this question, an analogy is provided here. It is known that in time-series analyses, the temporal autocorrelation of the residual of $\bar{Y}_t = \sum Y_{i,t}$ conditional on variables including variables that adjust for seasonality and long-term time-trend as unmeasured time-varying confounders can diminish and eventually randomly fluctuate around zero over time lag, which implies $Cov(\bar{Y}_t, \bar{Y}_{t-j} | z) \approx 0$ for $j > 0$. This is a desirable consequence of the adequate adjustment because the non-negligible crude autocorrelation, $Cov(\bar{Y}_t, \bar{Y}_{t-j}) \neq 0$ may be a sign of the existence of unmeasured time-varying confounding. It is also known that redundant adjustment for seasonality and long-term time-trend can make $Cov(\bar{Y}_t, \bar{Y}_{t-j} | z) < 0$. To apply analogy to our context, replace $Cov(\bar{Y}_t, \bar{Y}_{t-j} | z)$ with $Cov\left(x_{g,t-j}, x_{g,t-l} | x_g^{ep-l}, z\right)$. Recall that $x_g^{ep-l}$ may have time-trend (because $x_g^{-l}$ may have time-trend), for case-crossover designs, case-self-control matching scheme that can be seen as matching strata variables in $z$ is inherently related to time-trend (See Lu and Zeger 2007 for detail) and other covariates in $z$ that may also have time-trend. Thus, it would not be surprising that $Cov\left(x_{g,t-j}, x_{g,t-l} | x_g^{ep-l}, z\right) < 0$ can occur for some $j$s and its magnitude may be non-negligible. Finally, it is reasonable to consider not only

$$\eta_{j,l}^* \approx \frac{\eta_0 V(u_{g,t-j}^*)}{V\left(X_{g,t-l}^{ep}\Big| x_g^{ep-l}, z\right)}$$

but also

$$\eta_{j,l}^* \approx \frac{-\eta_0 V(u_{g,t-j}^*)}{V\left(X_{g,t-l}^{ep}\Big| x_g^{ep-l}, z\right)}$$

so that, by adopting $\pm$,

$$\theta_l^{L(L)} \approx R_{x_{g,t}^{ep}, x_g}^2 \left( \frac{\theta_l \pm \left(1 - R_{x_{g,t}^{ep}, x_g}^2\right) \sum_{j \neq l}^{L} \theta_j}{\eta_0} \right)$$

This approximation may perform well if residual confounding is not severe. This may perform in many types of errors because $u_{g,t-j}^*$ is designed from regression calibration (Web Appendix 3). For multiplicative L(L) errors, this approximation may be usable if treating this error as additive is reasonable and $R_{x_{g,t}^{ep}, x_g}^2$ and $\gamma_1^{crude}$ is known.

**Note.** This approximation relies on the assumption $R_{x_{g,t}^{ep}, x_g}^2 \approx R_{x_{g,t-l}^{ep}, x_{g,t-l} | x_g^{ep-l}, z}^2$. This assumption may not necessarily hold depending on, particularly $z$, which includes how to match case and self-controls. If $z$ may strongly predict $x_g$ (here "strong" is relative compared to the magnitude of the exposure measurement error) it may be likely $R_{x_{g,t}^{ep}, x_g}^2 \gg R_{x_{g,t-l}^{ep}, x_{g,t-l} | x_g^{ep-l}, z}^2$ because the correlation between the residual of $x_{g,t-l}^{ep}$ and the residual of $x_{g,t-l}$ would be mainly determined
53

by the exposure measurement error while the crude correlation between $x_{g,t-l}^{ep}$ and $x_{g,t-l}$ is deteremined by not just the error but also $z$.

**Web Appendix 9. Approximation for the estimator of a second-order polynomial.**

Non-linear terms such as splines and polynomials may be used. Considering exact variable transformation may be too tedious. It would be practical to focus on a second-order polynomial term, $x_{t-l}^{ep\ 2}$, because testing this term may demonstrate whether the linear term is inadequate to explain $Y$. Here, I do not use the subscript $i$ or $g$ for simplicity. $x_{t-l}^{ep}$ indicates an exposure variable having non-Berkson error.

If the exposure effect is believed to be non-linear only within the certain range of $x_{t-l}^{ep}$ (e.g., the effect of high temperature on a health outcome may manifest only after a certain threshold), the variable could be transformed to consider this.

Covariate adjustment should be considered as $R^2_{(x_{t-l}^{ep})^2 | z, z^2, x_{t-l}^{ep}, x_{t-l}^{ep} \times x^{ep-l}, x^{ep} \times z}$. $x_{t-l}^{ep} \times x^{ep-l}$ denotes interaction pairs of $x_{t-l}^{ep}$ and each element in $x^{ep-l}$. $z^2$ denote the squared variables in $z$. $x^{ep} \times z$ denotes interaction pairs of $x_{t-l}^{ep}$ and each element in $z$. The bias factor of the estimator for $(x_{t-l}^{ep})^2$ may be approximated as by assuming that residual confounding due to the use of $x^{ep-l}$ is negligible and the residual of $(x_{t-l}^{ep})^2$ conditional on $z, z^2, x_{t-l}^{ep}, x_{t-l}^{ep} \times x^{ep-l}, x^{ep} \times z$ and the residual of $x_{t-l}^{ep}$ conditional on $x^{ep-l} \times z, z, z^2$ follow a normal distribution, the bias factor becomes $\lambda_l^2$,

$$\lambda_l^2 \approx \frac{\left(R^2_{x_{t-l}^{ep}, x_{t-l} | x^{ep-l}, x^{ep-l} \times z, z, z^2}\right)^2}{\gamma_1^2} \approx \frac{\left(R^2_{x_t^{ep}, x_t}\right)^2}{\gamma_1^2}$$

Again, $R^2_{x_t^{ep},x_t}$ should be a good approximate. This bias factor and $\sigma^2_h(x_{t-l}^{ep\ 2})$ may be plugged into the equations developed eariler.

Let $\theta'_l$ be the coefficient for $(x_{t-l})^2$. If the residual of $(x_{t-l}^{ep})^2$ conditional on $\mathbf{z}, \mathbf{z}^2, x_{t-l}^{ep}, x_{t-l}^{ep} \times \mathbf{x}^{ep-l}, \mathbf{x}^{ep} \times \mathbf{z}$ and the residual of $x_{t-l}^{ep}$ conditional on $\mathbf{x}^{ep-l} \times \mathbf{z}, \mathbf{z}, \mathbf{z}^2$ approximately follow a normal distribution and residual confounding due to the use of $\mathbf{x}^{ep-l}$ is negligible, $\theta'^L_l$ in linear regression,

$$\theta'^L_l = \frac{Cov(Y, (x_{t-l}^{ep})^2 | \mathbf{z}, \mathbf{z}^2, x_{t-l}^{ep}, x_{t-l}^{ep} \times \mathbf{x}^{ep-l}, \mathbf{x}^{ep} \times \mathbf{z})}{V\left((x_{t-l}^{ep})^2 \big| \mathbf{z}, \mathbf{z}^2, x_{t-l}^{ep}, x_{t-l}^{ep} \times \mathbf{x}^{ep-l}, \mathbf{x}^{ep} \times \mathbf{z}\right)}$$

$$= \frac{Cov(Y, \gamma_0^2 + 2\gamma_0\gamma_1 x_{t-l} + 2u^L \gamma_1 x_{t-l} + 2\gamma_0 u_t^L + \gamma_1^2(x_{t-l})^2 + u_t^{L^2} | \mathbf{z}, \mathbf{z}^2, x_{t-l}^{ep}, x_{t-l}^{ep} \times \mathbf{x}^{ep-l}, \mathbf{x}^{ep} \times \mathbf{z})}{V\left((x_{t-l}^{ep})^2 \big| \mathbf{z}, \mathbf{z}^2, x_{t-l}^{ep}, x_{t-l}^{ep} \times \mathbf{x}^{ep-l}, \mathbf{x}^{ep} \times \mathbf{z}\right)}$$

$$\approx \frac{\left(2\gamma_0\gamma_1 Cov(Y, x_{t-l} | \mathbf{z}, \mathbf{z}^2, x_{t-l}^{ep}, x_{t-l}^{ep} \times \mathbf{x}^{ep-l}, \mathbf{x}^{ep} \times \mathbf{z}) + \gamma_1^2 Cov(Y, (x_{t-l})^2 | \mathbf{z}, \mathbf{z}^2, x_{t-l}^{ep}, x_{t-l}^{ep} \times \mathbf{x}^{ep-l}, \mathbf{x}^{ep} \times \mathbf{z})\right)}{2V(x_{t-l}^{ep} | \mathbf{x}^{ep-l}, \mathbf{x}^{ep-l} \times \mathbf{z}, \mathbf{z}, \mathbf{z}^2)^2}$$

$$\approx \frac{\frac{\left(\gamma_1^2 Cov(Y, (x_{t-l})^2 | \mathbf{z}, \mathbf{z}^2, x_{t-l}^{ep}, x_{t-l}^{ep} \times \mathbf{x}^{ep-l}, \mathbf{x}^{ep} \times \mathbf{z})\right)}{V((x_{t-l})^2 | \mathbf{z}, \mathbf{z}^2, x_{t-l}^{ep}, x_{t-l}^{ep} \times \mathbf{x}^{ep-l}, \mathbf{x}^{ep} \times \mathbf{z})} V((x_{t-l})^2 | \mathbf{z}, \mathbf{z}^2, x_{t-l}^{ep}, x_{t-l}^{ep} \times \mathbf{x}^{ep-l}, \mathbf{x}^{ep} \times \mathbf{z})}{2V(x_{t-l}^{ep} | \mathbf{x}^{ep-l}, \mathbf{x}^{ep-l} \times \mathbf{z}, \mathbf{z}, \mathbf{z}^2)^2}$$

$$= \frac{\gamma_1^2 \theta'_l V((x_{t-l})^2 | \mathbf{z}, \mathbf{z}^2, x_{t-l}^{ep}, x_{t-l}^{ep} \times \mathbf{x}^{ep-l}, \mathbf{x}^{ep} \times \mathbf{z})}{2V(x_{t-l}^{ep} | \mathbf{x}^{ep-l}, \mathbf{x}^{ep-l} \times \mathbf{z}, \mathbf{z}, \mathbf{z}^2)^2} \approx \frac{2V\left(x_{t-l} | x_t^{ep-l}, x_t^{ep-l} \times \mathbf{z}, \mathbf{z}, \mathbf{z}^2\right)^2 \gamma_1^2 \theta'_l}{2V(x_{t-l}^{ep} | \mathbf{x}^{ep-l}, \mathbf{x}^{ep-l} \times \mathbf{z}, \mathbf{z}, \mathbf{z}^2)^2}$$

$$= \frac{\gamma_1^2 V(x_{t-l} | \mathbf{x}^{ep-l}, \mathbf{x}^{ep-l} \times \mathbf{z}, \mathbf{z}, \mathbf{z}^2)^2}{V(x_{t-l}^{ep} | \mathbf{x}^{ep-l}, \mathbf{x}^{ep-l} \times \mathbf{z}, \mathbf{z}, \mathbf{z}^2)^2} \theta'_l = \lambda_l^2 \theta'_l$$

where $\lambda_l^2 = \left(\frac{\gamma_1^2 V(x_{t-l} | \mathbf{x}^{ep-l}, \mathbf{x}^{ep-l} \times \mathbf{z}, \mathbf{z}, \mathbf{z}^2)}{V(x_{t-l}^{ep} | \mathbf{x}^{ep-l}, \mathbf{x}^{ep-l} \times \mathbf{z}, \mathbf{z}, \mathbf{z}^2)}\right)^2 / \gamma_1^2 = \frac{\left(R^2_{x_{t-l}^{ep}, x_{t-l} | \mathbf{x}^{ep-l}, \mathbf{x}^{ep-l} \times \mathbf{z}, \mathbf{z}, \mathbf{z}^2}\right)^2}{\gamma_1^2} \approx \frac{\left(R^2_{x_t^{ep}, x_t}\right)^2}{\gamma_1^2}$

This may be applicable for non-linear models such as logistic regression when the effect size is small, or the measurement error is small.

**Web Appendix 10. Simulation experiments.**

Simulation experiments were motivated by air pollution modelled estimates. Web Figure 1 presents relationships of ground-based measurements for daily $O_3$ and $PM_{2.5}$ with estimated levels in Rio de Janeiro, Brazil (2010–2017) (A, B, E, and F) obtained from elsewhere (Kim et al. , 2024) and estimated levels in the contiguous United States (2020 summer) (C, D, G, and H) from a frequently cited model in air pollution epidemiology (Reff, 2023) (i.e., Modelled estimates of $PM_{2.5}$, $O_3$, (United States Environmental Protection Agency (EPA)'s Fused Air Quality Surface Using Downscaling (FAQSD)). These relationships were identified using linear regression (red lines) from the BL error perspective (A–D) and from the LL error perspective. Depending on the prediction modeling strategy, the correct perspective in health effect estimation may differ, which is beyond the scope of this paper. Web Table 1 show findings of linear regressions for data in the United States suggesting the error of these estimates may be BL or LL error in health effect estimation. For all census tracts in the United States for June to August 2020 from the United States, a generalized additive model was fit to regress the difference between the daily FAQSD $PM_{2.5}$ estimate (Lag0) and daily $PM_{2.5}$ measurement (Lag0) at EPA monitor against census tract-level FAQSD $PM_{2.5}$ estimates from (Lag0, Lag1, and Lag2) and a three-way spline (Longitude, Latitude, Time). Generalized cross-validation was used to determine degrees of freedom of the spline. All terms were statistically significant, but the model explained the dependent variable ≤10% according to the deviance explained. This low deviance explained implies the existence of unexplained variability of the difference to a great degree. The residuals were still not randomly distributed. Web Figure 2 show residual plots.

Simulation experiments were conducted. These mimicked identifications of the association between short-term exposure to PM$_{2.5}$ and a health outcome, using PM$_{2.5}$ variables measured with errors. Confounders to be considered include short-term exposure to O$_3$ and temperature, as well as unmeasured time-varying confounders that manifest as seasonality, time-trend, and day-of-the-week effects. Several additive and multiplicative BL, and LL errors in PM$_{2.5}$ variables were considered, using ground-based measurements at EPA monitors and modelled estimates. Web Figure 3 present scatter plots for gold standard variables and variables measured with error in simulations.

Specifically, the predicted PM$_{2.5}$ variable was obtained from the fitted generalized additive model for all census tracts in Chicago and then was normalized. Then, the covariance of this predicted variable (i.e., spatiotemporal covariance) was non-parametrically estimated using the method developed by Yang and Qiu (2019). Several exposure measurement error scenarios for linear-like errors were considered.

$X_{i,t}^{ep} = \gamma_1 X_{i,t} + \kappa_{additive} \times error_{1,i,t} + \gamma_0$ for additive linear-like errors

$X_{i,t}^{ep} = \exp(\gamma_1 \log(X_{i,t}) + \kappa_{multiplicative} \times error_{1,i,t} - error_{2,i,t}/2)$ for multiplicative linear-like errors, where $error_{1,i,t}$ follows a multivariate normal distribution with the covariance estimated and $error_{2,i,t}$ follows a gamma distribution. $X_{i,t}$ is a FAQSD estimate, which is spatiotemporally autocorrelated. $error_{1,i,t}$ is highly spatiotemporally autocorrelated. $error_{2,i,t}$ is completely random. Several $\kappa$ values were used. A large $\kappa$ is reasonable to mimick real-world scenarios because the generalized additive model used above was able to explain the difference to a very limited degree.

For Berkson-like errors,

$$X_{i,t}^{new} = \exp(1.2X_{i,t} + 7error_{1,i,t} - error_{2,i,t})$$

and then, $X_{g,t}$ was created using a population-weighted average, where $g$ represents ZIP Code in Chicago. $E_g[X_{i,t}^{new}]$ was used for additive Berkson-like error. $\exp(E_g[\log(X_{i,t}^{new})])$ was used for multiplicative Berkson-like error. The US Department of Housing and Urban Development's USPS ZIP Code-Census Tract Crosswalk file was used to create population-weighted averages at each ZIP Code. A total of 784 census tracts and 89 ZIP Codes in Chicago were considered.

The outcome generating model was designed to draw samples using $Y_{i,t} \sim Binom(p_{i,t})$ where $p_{i,t} = Kp_{0,i,t}\exp\left(\sum_{l=0}^{2}\theta_l PM2.5_{i,t-l} + \sum_{l=0}^{2}\vartheta_l O3_{i,t-l} + \sum_{l=0}^{1}\varsigma_l HighTemp_{i,t-l}\right)$. $p_{0,i,t}$, time-varying baseline hazard, was determined using statistics on underlying cause of death and population size from the Centers for Disease Control and Prevention's WONDER and 2020 decennial Census from the Census Bureau:

$\theta_0 = 0.001$, $\theta_1 = 0.0024$, and $\theta_3 = 0.0006$

$\vartheta_0 = 0.0001$, $\vartheta_1 = 0.0003$, and $\vartheta_2 = 0.0001$

$\varsigma_0 = 0.01$ and $\varsigma_1 = 0.005$

PM$_{2.5}$ level ranged from 0.24 µg/m³ to 163.5 µg/m³ with the variance of 13.8µg/m³. O$_3$ level ranged from 5.6 ppb to 122.6 ppb with the variance of 164.6 ppb. The range for high temperature is from 0 to 14.6°C. Temperature data was obtained from *Daymet* and O$_3$ data was obtained from FAQSD.

For each simulation sample, time-stratified case-crossover analysis with conditional logistic regression was conducted to estimate $\theta_l$. The percentage of rejecting the null hypothesis (empirical statistical power), $E[\hat{\theta}_l]$ and $E\left[\widehat{SE}[\hat{\theta}_l]\right]$ were obtained.

$n$, $E[\hat{\theta}_l]$, $E\left[\hat{\bar{\theta}}\right]$, $E[\widehat{SE}(\hat{\theta}_l)]$, and $E\left[\widehat{SE}\left(\hat{\bar{\theta}}\right)\right]$ were also calculated using developed approximation equations earlier. They were compared to corresponding empirical quantities estimated from time-stratified case-crossover analyses for 2,500 simulation runs for each error scenario. Findings demonstrate that the calculations perform well. Web Tables 2 and 3 demonstrate empirical effect estimates and standard error estimates are consistent with the approximates. As expected, Eqs.19 and 20 may sometimes provide accurate approximates for effect estimates, suggesting that it would not be easy to accurately consider bias in the estimator without validation data when residual confounding due to distributed lags measured with error is not small. Web Figure 4 shows the consistency between empirical statistical powers (i.e., the percentage of rejecting the null hypothesis when the null hypothesis is false) and the calculated power curves. The inconsistencies were found in only very low statistical power settings for $\theta_l$ estimation. These may have been attributed to a failure of the convergence due to correlation between distributed lags in such settings because the accuracy (See "Approx. 1" in Web Table 2) and precision of the estimators were still accurately calculated (Web Table 3).

**Web Table 1. Linear regressions of measured O₃/PM₂.₅ at EPA monitors against distributed temporal lags of estimated O₃/PM₂.₅ (Berkson-like error perspective) by U.S. EPA's FAQSD models (Jun–Aug. 2020) and of estimated O₃/PM₂.₅ by U.S. EPA's FAQSD models (Jun–Aug. 2020) against distributed temporal lags of measured O₃/PM₂.₅ at EPA monitors (Linear-like error perspective)**

**Note.** Web Figure 1 presents residual plots including partial autocorrelation plots that demonstrate that residuals are not random.

| Parameter | Regression coefficient estimates (95% confidence intervals) | | | |
|---|---|---|---|---|
| | $O_3$ | | $PM_{2.5}$ | |
| | Berkson-Like error perspective | Linear-Like error perspective | Berkson-Like error perspective | Linear-Like error perspective |
| Intercept | -0.931 (-1.011, -0.851) | 2.225 (2.152, 2.299) | -0.362 (-0.404, -0.321) | 0.901 (0.866, 0.936) |
| Lag0 | 1.029 (1.027, 1.032) | 0.929 (0.927, 0.931) | 1.065 (1.061, 1.070) | 0.840 (0.837, 0.844) |
| Lag1 | -0.009 (-0.011, -0.006) | 0.023 (0.021, 0.025) | -0.026 (-0.032, -0.020) | 0.059 (0.055, 0.063) |
| Lag2 | -0.001 (-0.004, 0.001) | 0.002 (0, 0.004) | 0 (-0.006, 0.005) | 0 (-0.005, 0.004) |
| Lag3 | 0 (-0.002, 0.003) | 0.001 (-0.001, 0.003) | -0.006 (-0.012, -0.001) | 0.008 (0.003, 0.012) |
| Lag4 | 0 (-0.002, 0.003) | -0.002 (-0.004, 0) | 0.002 (-0.004, 0.008) | -0.003 (-0.007, 0.001) |
| Lag5 | 0.001 (-0.001, 0.004) | -0.003 (-0.005, -0.001) | 0 (-0.006, 0.006) | 0 (-0.004, 0.004) |
| Lag6 | 0 (-0.002, 0.003) | -0.002 (-0.004, 0) | 0.001 (-0.005, 0.006) | -0.002 (-0.006, 0.002) |
| Lag7 | 0 (-0.002, 0.002) | -0.002 (-0.003, 0) | -0.005 (-0.01, -0.001) | 0.003 (-0.001, 0.006) |
| R-sq. | 96.6% | 96.6% | 92.0% | 92.2% |

**Web Table 2. Empirical estimates of $\theta_l$ and $\bar{\theta}$ and various approximations for $\theta_l$ and $\bar{\theta}$ in simulation experiments**

**Note.** MLL=Multiplicative Linear-Like error; ALL=Additive Linear-Like error; ABL=Additive Berkson-Like error; MBL=Multiplicative Berkson-Like error. See Figure 3 for distributions of MLL1–4, ALL1–2, ABL, and MBL.

Lag0=$\theta_0$; Lag1=$\theta_1$; Lag2=$\theta_2$; and Lag02=$\bar{\theta} = \theta_0 + \theta_1 + \theta_2$. $\theta_0 = 0.001$, $\theta_1 = 0.0024$, $\theta_3 = 0.0006$, and $\bar{\theta} = 0.004$ were used to generate simulation samples. Except for ABL, most of the effect estimates were biased due to the error. See "a. Empirical", which shows $E[\hat{\theta}_l] \times 1000$.

| Error | Quantities | Lag0 | Lag1 | Lag2 | Lag02 |
|---|---|---|---|---|---|
| No error | $\theta$ used to generate data ×1000 | 1.00 | 2.40 | 0.60 | 4.00 |
| MLL1 | a. Empirical (×1000) | 1.62 | 3.38 | 1.02 | 6.01 |
| | b. Approx.1 (×1000) [Web Appendix 3] | 1.56 | 3.42 | 1.01 | 5.99 |
| | c. Difference (%) (i.e., (b-a)/a) | -3.4 | 1.2 | -0.7 | -0.4 |
| | d. Approx.2 (×1000) [Eq. 19] "+" | 1.71 | 3.73 | 1.13 | 6.57 |
| | e. Difference (%) (i.e., (d-a)/a) | 5.8 | 10.5 | 11.3 | 9.4 |
| | f. Approx.3(×1000) [Eq. 19] "-" | 1.31 | 3.52 | 0.68 | 5.52 |
| | g. Difference (%) (i.e., (f-a)/a) | -18.7 | 4.3 | -32.8 | -8.2 |
| | h. Approx.4 (×1000) [Eq. 20] | 1.51 | 3.63 | 0.91 | 6.05 |
| | i. Difference (%) (i.e., (h-a)/a) | -6.4 | 7.4 | -10.7 | 0.6 |
| MLL2 | a. Empirical (×1000) | 1.08 | 3.08 | 0.64 | 4.81 |
| | b. Approx.1 (×1000) [Web Appendix 3] | 1.09 | 3.14 | 0.64 | 4.87 |
| | c. Difference (%) (i.e., (b-a)/a) | 0.9 | 1.7 | -0.2 | 1.3 |
| | d. Approx.2 (×1000) [Eq. 19] "+" | 1.8 | 3.65 | 1.27 | 6.73 |
| | e. Difference (%) (i.e., (d-a)/a) | 67.2 | 18.5 | 98.4 | 40.1 |
| | f. Approx.3(×1000) [Eq. 19] "-" | 1.07 | 3.26 | 0.45 | 4.78 |
| | g. Difference (%) (i.e., (f-a)/a) | -0.7 | 5.8 | -30.2 | -0.5 |
| | h. Approx.4 (×1000) [Eq. 20] | 1.44 | 3.46 | 0.87 | 5.77 |
| | i. Difference (%) (i.e., (h-a)/a) | 33.7 | 12.3 | 34.9 | 20.1 |
| MLL3 | a. Empirical (×1000) | 0.79 | 1.76 | 0.50 | 3.05 |
| | b. Approx.1 (×1000) [Web Appendix 3] | 0.77 | 1.80 | 0.48 | 3.05 |
| | c. Difference (%) (i.e., (b-a)/a) | -2.2 | 2.2 | -3.0 | 0.2 |
| | d. Approx.2 (×1000) [Eq. 19] "+" | 0.86 | 1.94 | 0.56 | 3.36 |
| | e. Difference (%) (i.e., (d-a)/a) | 9.5 | 10.1 | 11.8 | 10.2 |
| | f. Approx.3(×1000) [Eq. 19] "-" | 0.72 | 1.86 | 0.39 | 2.96 |
| | g. Difference (%) (i.e., (f-a)/a) | -9.3 | 5.6 | -21.8 | -2.7 |
| | h. Approx.4 (×1000) [Eq. 20] | 0.79 | 1.90 | 0.47 | 3.16 |
| | i. Difference (%) (i.e., (h-a)/a) | 0.1 | 7.9 | -5.0 | 3.7 |
| MLL4 | a. Empirical (×1000) | 0.58 | 1.67 | 0.36 | 2.61 |
| | b. Approx.1 (×1000) [Web Appendix 3] | 0.60 | 1.70 | 0.35 | 2.64 |
| | c. Difference (%) (i.e., (b-a)/a) | 2.4 | 1.4 | -3.1 | 1.0 |
| | d. Approx.2 (×1000) [Eq. 19] "+" | 0.90 | 1.91 | 0.62 | 3.43 |
| | e. Difference (%) (i.e., (d-a)/a) | 55.6 | 13.9 | 73.3 | 31.3 |
| | f. Approx.3 (×1000) [Eq. 19] "-" | 0.62 | 1.75 | 0.29 | 2.67 |
| | g. Difference (%) (i.e., (g-a)/a) | 6.4 | 4.9 | -17.4 | 2.2 |
| | h. Approx.4 (×1000) [Eq. 20] | 0.76 | 1.83 | 0.46 | 3.05 |
| | i. Difference (%) (i.e., (h-a)/a) | 31.2 | 9.5 | 28.4 | 16.9 |
| ALL1 | a. Empirical (×1000) | 0.13 | 2.19 | -0.15 | 2.16 |
| | b. Approx.1 (×1000) [Web Appendix 3] | 0.11 | 2.17 | -0.14 | 2.15 |
| | c. Difference (%) (i.e., (b-a)/a) | -10.9 | -0.8 | -9.8 | -0.7 |
| | d. Approx.2 (×1000) [Eq. 19] "+" | 1.35 | 2.48 | 1.04 | 4.88 |
| | e. Difference (%) (i.e., (d-a)/a) | 957.8 | 13.4 | -780.4 | 125.5 |
| | f. Approx.3 (×1000) [Eq. 19] "-" | 0.49 | 2.00 | 0.06 | 2.55 |
| | g. Difference (%) (i.e., (f-a)/a) | 282.0 | -8.7 | -140.5 | 17.9 |
| | h. Approx.4 (×1000) [Eq. 20] | 0.93 | 2.24 | 0.56 | 3.74 |
| | i. Difference (%) (i.e., (h-a)/a) | 629.7 | 2.5 | -466.0 | 72.8 |
| ALL2 | a. Empirical (×1000) | -0.44 | 1.90 | -0.57 | 0.90 |
| | b. Approx.1 (×1000) [Web Appendix 3] | -0.45 | 1.87 | -0.60 | 0.82 |
| | c. Difference (%) (i.e., (b-a)/a) | 3.4 | -1.8 | 4.5 | -8.5 |
| | d. Approx.2 (×1000) [Eq. 19] "+" | 1.36 | 1.92 | 1.21 | 4.49 |
| | e. Difference (%) (i.e., (d-a)/a) | -413.3 | 0.8 | -310.6 | 400.4 |
| | f. Approx.3 (×1000) [Eq. 19] "-" | -0.16 | 1.07 | -0.50 | 0.42 |
| | g. Difference (%) (i.e., (f-a)/a) | -63.4 | -43.6 | -13.3 | -53.4 |
| | h. Approx.4 (×1000) [Eq. 20] | 0.63 | 1.50 | 0.38 | 2.50 |
| | i. Difference (%) (i.e., (h-a)/a) | -243.7 | -21.2 | -165.4 | 178.7 |
| ABL | a. Empirical (×1000) | 1.07 | 2.38 | 0.61 | 4.05 |
| | b. Approx.1 (×1000) [Web Appendix 3] | 1.02 | 2.45 | 0.61 | 4.08 |
| | c. Difference (%) (i.e., (b-a)/a) | -4.7 | 3.0 | 1.5 | 0.7 |
| | d. Approx.3 (×1000) [Eq. 20] | 1.02 | 2.44 | 0.61 | 4.07 |
| | e. Difference (%) (i.e., (d-a)/a) | -4.6 | 2.7 | 1.0 | 0.5 |
| MBL | a. Empirical (×1000) | 1.08 | 3.02 | 0.61 | 4.71 |
| | b. Approx.1 (×1000) [Web Appendix 3] | 1.14 | 2.84 | 0.66 | 4.64 |
| | c. Difference (%) (i.e., (b-a)/a) | 5.1 | -5.9 | 9.0 | -1.4 |
| | d. Approx.3 (×1000) [Eq. 20] | 1.13 | 2.71 | 0.68 | 4.52 |
| | e. Difference (%) (i.e., (d-a)/a) | 4.6 | -10.1 | 11.5 | -3.9 |

**Web Table 3. Empirical standard error estimates and approximations in simulation experiments**

**Note.** Lag0=$\theta_0$; Lag1=$\theta_1$; Lag2=$\theta_2$; and Lag02=$\bar{\theta} = \theta_0 + \theta_1 + \theta_2$
MLL=Multiplicative Linear-Like error; ALL=Additive Linear-Like error; ABL=Additive Berkson-Like error; MBL=Multiplicative Berkson-Like error. See Figure 3 for distributions of MLL1–4, ALL1–2, ABL, and MBL.

| Error | Quantities | Lag0 | Lag1 | Lag2 | Lag02 |
|---|---|---|---|---|---|
| MLL1 | a. Empirical (×1000) | 1.535 | 1.648 | 1.489 | 2.007 |
|  | b. Approx. (×1000) (Eqs. 15 and 16) | 1.491 | 1.611 | 1.449 | 1.993 |
|  | c. Difference (%) (i.e., (b-a)/a) | -2.8 | -2.3 | -2.7 | -0.7 |
| MLL2 | a. Empirical (×1000) | 1.478 | 1.626 | 1.447 | 1.813 |
|  | b. Approx. (×1000) (Eqs. 15 and 16) | 1.473 | 1.617 | 1.441 | 1.783 |
|  | c. Difference (%) (i.e., (b-a)/a) | -0.3 | -0.5 | -0.4 | -1.7 |
| MLL3 | a. Empirical (×1000) | 0.797 | 0.869 | 0.778 | 1.016 |
|  | b. Approx. (×1000) (Eqs. 15 and 16) | 0.768 | 0.843 | 0.753 | 0.999 |
|  | c. Difference (%) (i.e., (b-a)/a) | -3.7 | -2.9 | -3.2 | -1.6 |
| MLL4 | a. Empirical (×1000) | 0.774 | 0.859 | 0.760 | 0.942 |
|  | b. Approx. (×1000) (Eqs. 15 and 16) | 0.761 | 0.844 | 0.749 | 0.917 |
|  | c. Difference (%) (i.e., (b-a)/a) | -1.7 | -1.7 | -1.4 | -2.7 |
| ALL1 | a. Empirical (×1000) | 1.041 | 1.247 | 1.033 | 1.037 |
|  | b. Approx. (×1000) (Eqs. 12 and 14) | 1.119 | 1.332 | 1.109 | 1.006 |
|  | c. Difference (%) (i.e., (b-a)/a) | 7.5 | 6.8 | 7.3 | -3.0 |
| ALL2 | a. Empirical (×1000) | 0.917 | 1.234 | 0.928 | 0.656 |
|  | b. Approx. (×1000) (Eqs. 12 and 14) | 0.957 | 1.292 | 0.969 | 0.644 |
|  | c. Difference (%) (i.e., (b-a)/a) | 4.4 | 4.7 | 4.4 | -1.9 |
| ABL | a. Empirical (×1000) | 4.386 | 6.794 | 4.547 | 2.267 |
|  | b. Approx. (×1000) (Eqs. 8 and 10) | 4.418 | 6.774 | 4.575 | 2.239 |
|  | c. Difference (%) (i.e., (b-a)/a) | 0.7 | -0.3 | 0.6 | -1.2 |
| MBL | a. Empirical (×1000) | 5.568 | 8.828 | 5.777 | 2.657 |
|  | b. Approx. (×1000) (Eqs. 8 and 10) | 5.773 | 9.074 | 5.956 | 2.673 |
|  | c. Difference (%) (i.e., (b-a)/a) | 3.7 | 2.8 | 3.1 | 0.6 |

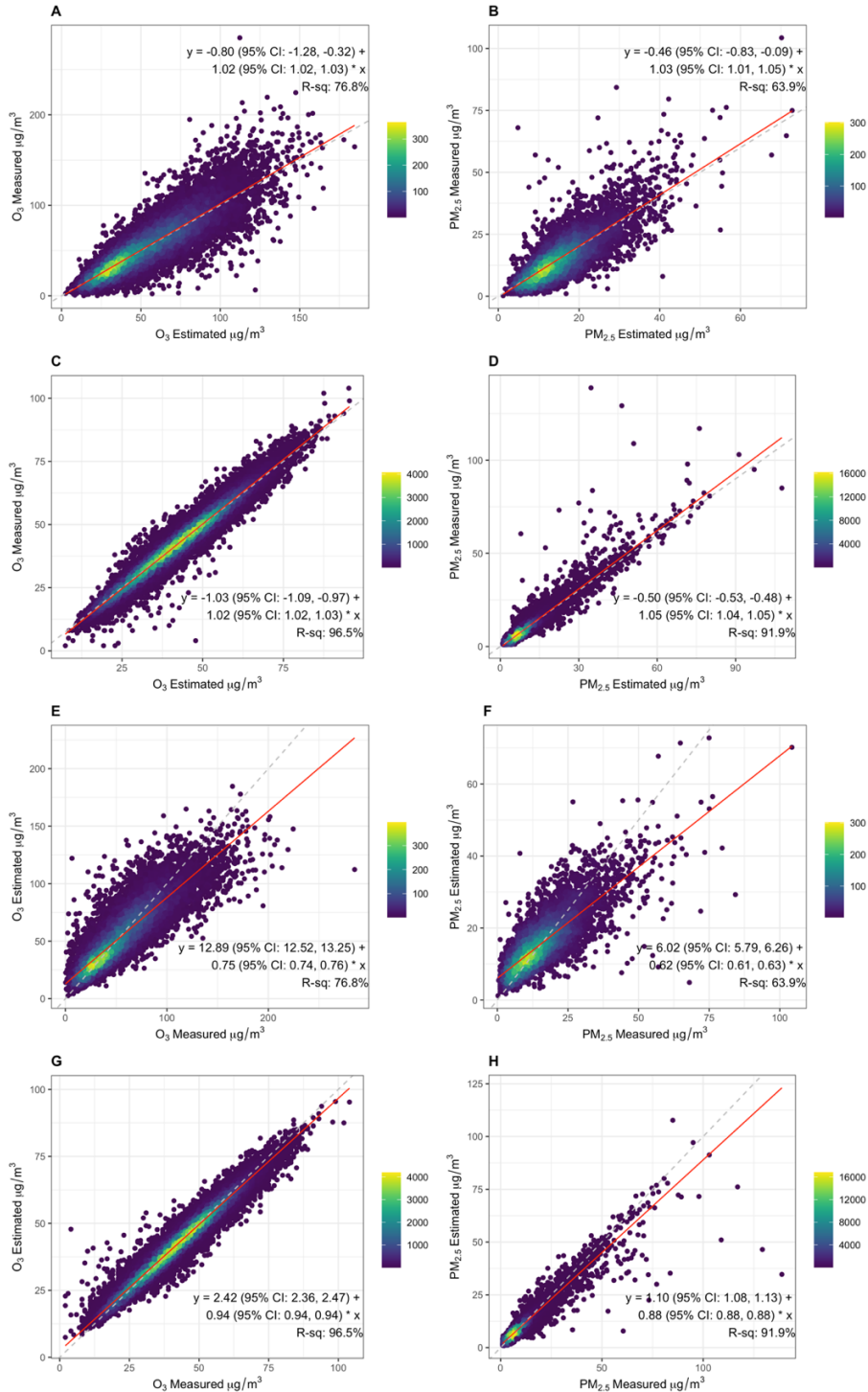

**Web Figure 1. Relationship between daily estimated $O_3$/$PM_{2.5}$ and daily measured $O_3$/$PM_{2.5}$ at monitors in Rio de Janeiro, Brazil for the years 2012–2017 (A, B, E, and F) and in the contiguous United States (C, D, G, and H) for Jun.–Aug. 2020**

Note. The difference between the top four panels (Berkson-like error perspective, A–D) and the bottom four panels (linear-like error perspective, E–H) is that the same data were analyzed differently by switching the x and y axes. Dashed diagonal lines indicates 100% agreement. Red solid lines indicate linear regression fit. Points indicate the number of data points.

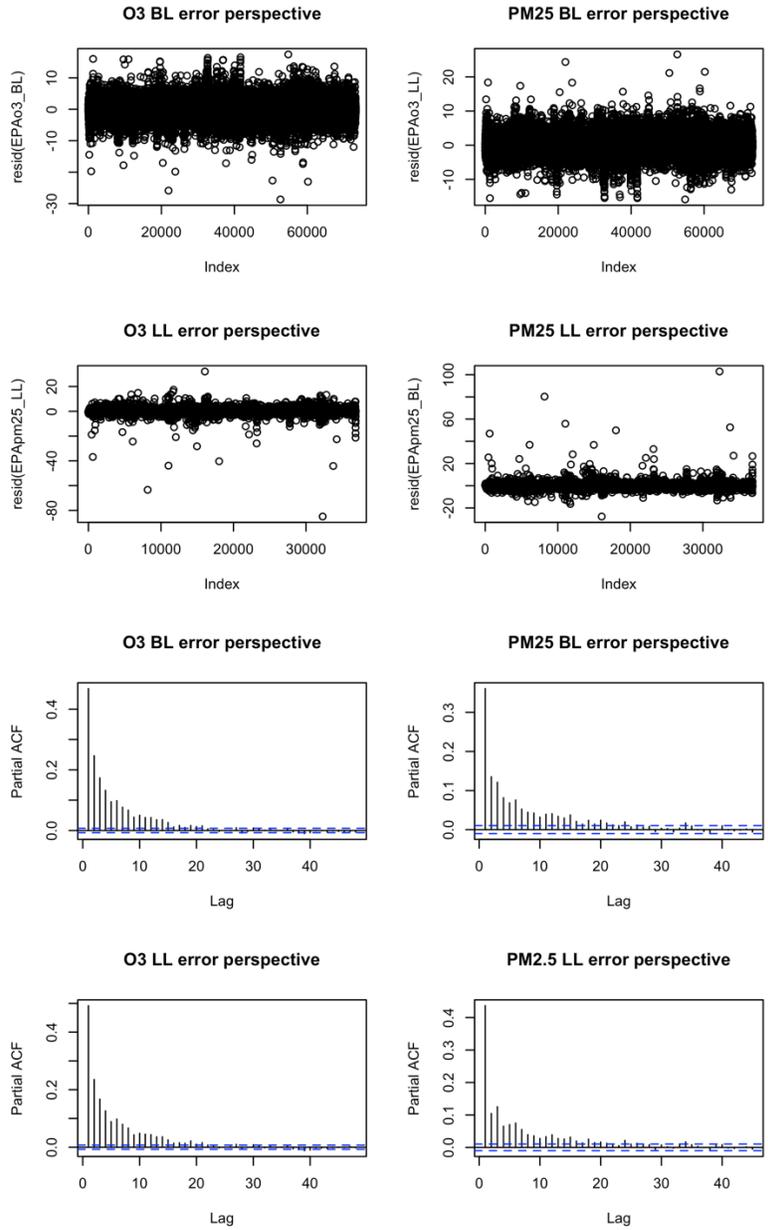

**Web Figure 2. Residual plots including partial autocorrelation plots for linear regressions in Table 1 of the main manuscript**

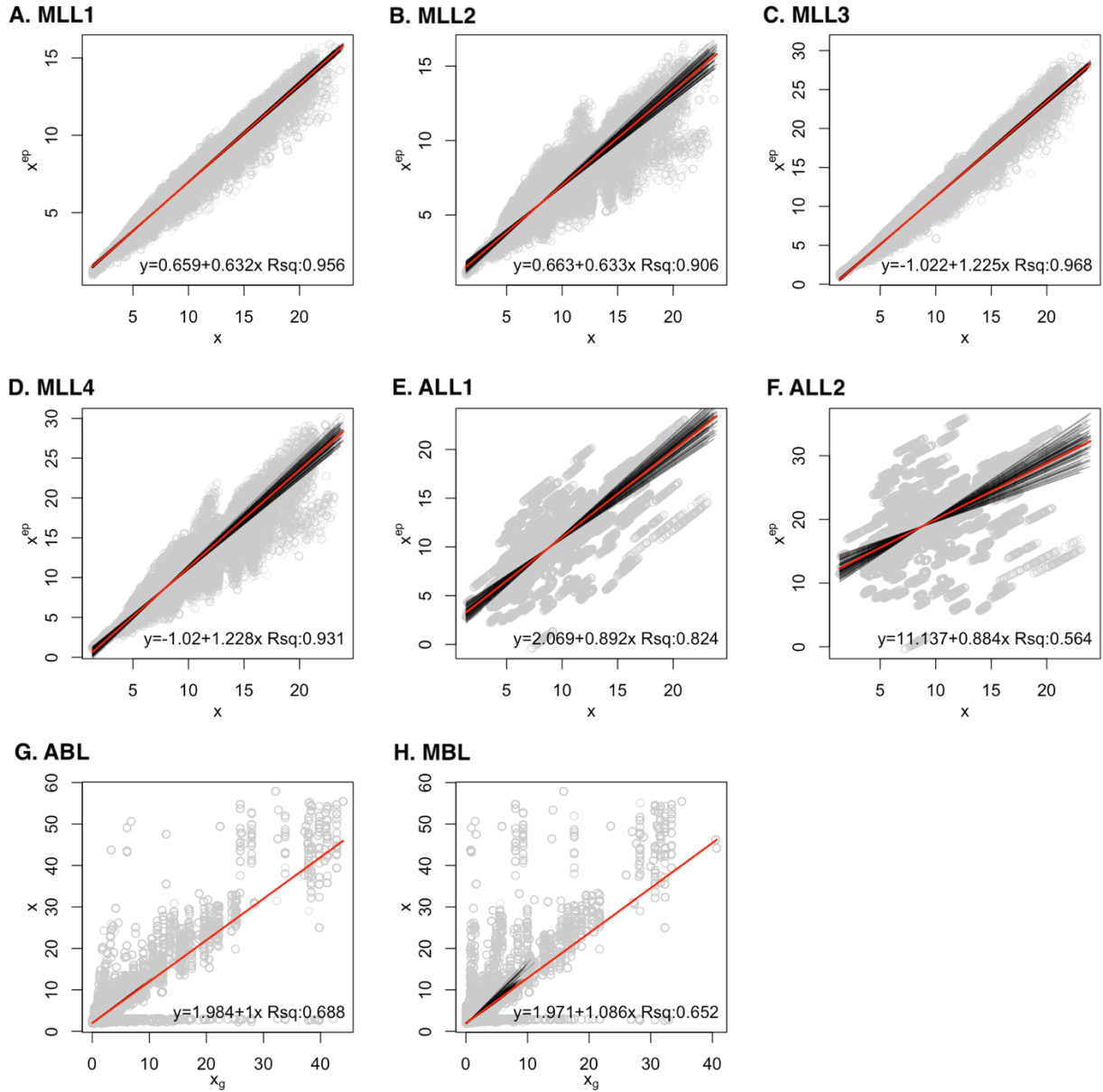

**Web Figure 3. Scatter plots for several sets of gold standard variables and variables measured with error created in simulation experiments**

**Note.** Eight measurement error scenarios were created. MLL=Multiplicative Linear-Like error; ALL=Additive Linear-Like error; ABL=Additive Berkson-Like error; and MBL=Multiplicative Berkson-Like error. Only randomly selected 10,000 data points from each of the first 100 simulation samples (for MLL and ALL) and randomly selected 10,000 data points from each of the first 250 simulation samples (for ABL and MBL) are presented to enhance visualizations. Red solid lines represent the linear regression fit for each sample that consists of 10,000 data points and represent the average of these linear regression fits. For MLL data generation, $\gamma_1$=0.9 (MLL1 and MLL2) or $\gamma_1$=1.1 (MLL3 and MLL4) were used.

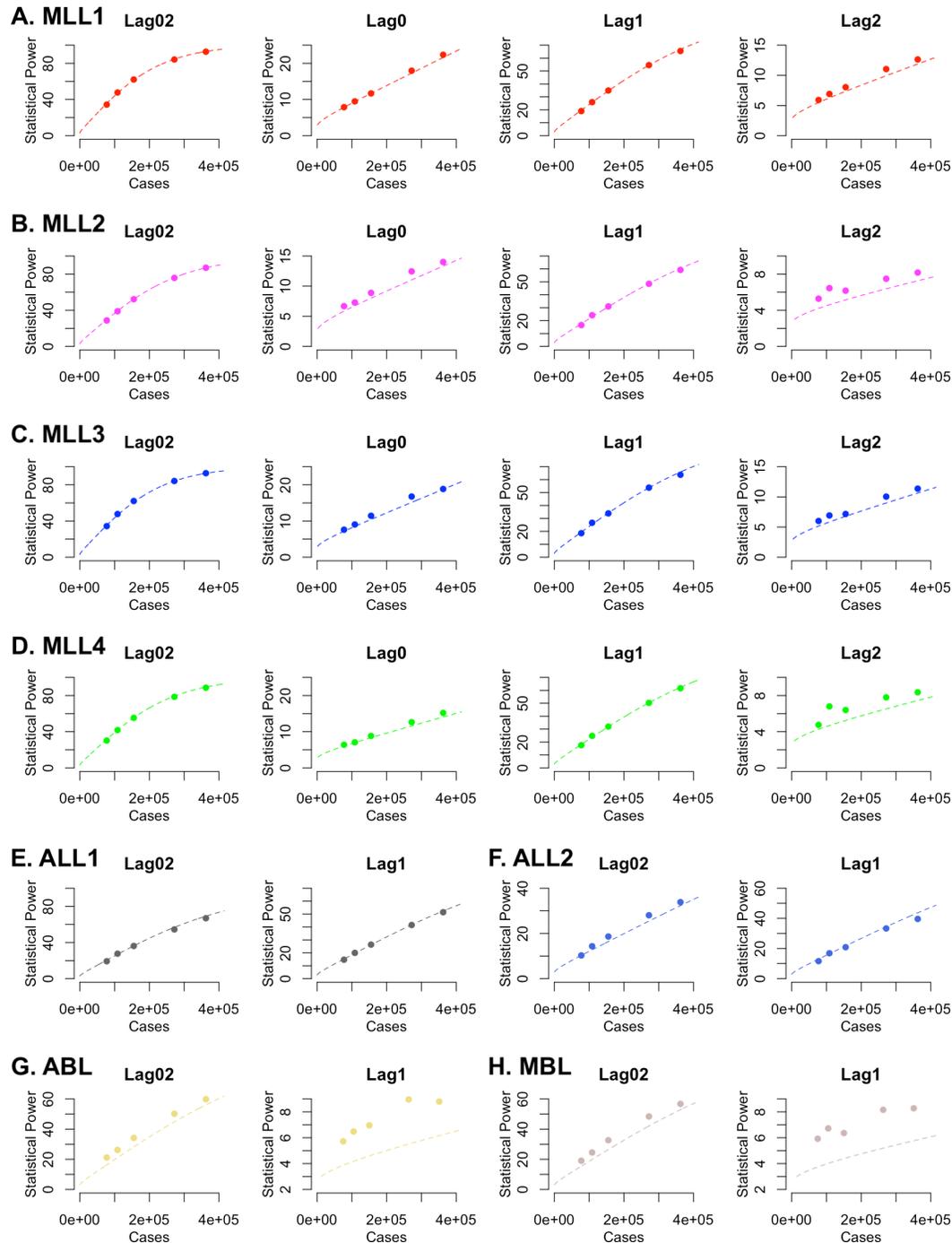

**Web Figure 4. Power curves obtained from sample size calculation equations presented in the main text (dashed curves) and empirical statistical powers (points) from simulation experiments**

**Note.** Each data point was obtained from 2,500 simulation samples. Different exposure measurement error scenarios were considered: MLL (Multiplicative Linear Like Error); ALL (Additive Linear Like Error); ABL (Additive Berkson Like Error); MBL (Multiplicative Berkson Like Error). For ALL1, ALL2, ABL, and MBL, lag0 and lag2 were not presented here because empirical statistical powers were extremely low.

**References in Online Supplementary Materials**

Yang K, Qiu P. Nonparametric estimation of the spatio-temporal covariance structure. *Statistics in Medicine* 2019;**38**(23):4555-4565.